\documentclass[%
 reprint,
superscriptaddress,
nofootinbib,
 amsmath,amssymb,
aps,
twocolumn,
prd,
]{revtex4-2}

\usepackage{graphicx}
\usepackage{dcolumn}
\usepackage{bm}
\usepackage[colorlinks]{hyperref}

\def\mnras{\rm{MNRAS}}
\def\aj{\rm{AJ}}

\def\aap{\rm{A\&A}}

\def\apj{\rm{ApJ}}                 
\def\jcap{\rm{JCAP}}

\usepackage{color}
\usepackage{xspace}
\usepackage[caption=false]{subfig}
\def\beq{\begin{eqnarray}}
\def\eeq{\end{eqnarray}}

\usepackage[normalem]{ulem}
\usepackage{physics}
\usepackage{ifthen}
\usepackage{dcolumn}

\graphicspath{{figures/}}

\newcommand{\Mpch}{h^{-1}\mathrm{Mpc}}
\newcommand{\hMpc}{h\,\mathrm{Mpc}^{-1}}
\newcommand{\hun}{\,\mathrm{km}\,\mathrm{s}^{-1}\mathrm{Mpc}^{-1}}
\newcommand{\lcdm}{$\Lambda$CDM\xspace}

\newcommand{\dve}[4][1]{
\ifthenelse{\equal{#1}{1}}{\left.\dv{#2}{#3}\right\vert_{#4}}{\left.\dv[#1]{#2}{#3}\right\vert_{#4}}}

\newcommand{\pdve}[4][1]{
	\ifthenelse{\equal{#1}{1}}{\left.\pdv{#2}{#3}\right\vert_{#4}}{\left.\pdv[#1]{#2}{#3}\right\vert_{#4}}}

\definecolor{darkgreen}{RGB}{0,120,0}
\definecolor{brown}{RGB}{120,60,0}

\newcommand{\resub}[1]{#1}

\begin{document}


\title{Determining the Hubble constant without the sound horizon: Perspectives with future galaxy surveys}
\author{Gerrit S. Farren}%
\affiliation{Department of Applied Mathematics and Theoretical Physics, University of Cambridge,\\ Cambridge CB3 0WA, UK}
\author{Oliver H.\,E. Philcox}
\affiliation{Department of Astrophysical Sciences, Princeton University,\\ Princeton, NJ 08540, USA}
\affiliation{School of Natural Sciences, Institute for Advanced Study, 1 Einstein Drive,\\ Princeton, NJ 08540, USA}
\author{Blake D. Sherwin}%
\affiliation{Department of Applied Mathematics and Theoretical Physics, University of Cambridge,\\ Cambridge CB3 0WA, UK}
\affiliation{Kavli Institute for Cosmology, Institute of Astronomy, University of Cambridge,\\ Cambridge CB3 0HA, UK}%



\begin{abstract} 
$H_0$ constraints from galaxy surveys are sourced by the geometric properties of two standardisable rulers: the sound horizon scale, $r_s$, and the matter-radiation equality scale, $k_{\rm eq}$. While most analyses over the past decade have focused on the first scale, recent work has emphasised that the second can provide an independent source of information about the expansion rate of the universe. In this work, we demonstrate an improved method for performing such a measurement with future galaxy surveys such as Euclid. Previous approaches have avoided $r_s$-based information by removing the prior on the baryon density, and thus the sound-horizon calibration. Here, we present a new method to marginalise over $r_s$; this allows baryon information to be retained, which enables tighter parameter constraints. For a Euclid-like spectroscopic survey, we forecast sound-horizon independent $H_0$ constraints of $\sigma_{H_0} = 0.7\,\hun$ for our method using the equality scale, compared with $\sigma_{H_0} = 0.5\,\hun$ from the sound horizon. Upcoming equality scale $H_0$ measurements thus can be highly competitive, although we caution that the impact of observational systematics on such measurements still needs to be investigated in detail. Applying our new approach to the BOSS power spectrum gives $H_0 = 69.5^{+3.0}_{-3.5}\hun$ from equality alone, somewhat tighter than previous constraints. Consistency of $r_s$- and $k_{\rm eq}$-based $H_0$ measurements can provide a valuable internal consistency test of the cosmological model; as an example, we consider the change in $H_0$ created by early dark energy. Assuming the \textit{Planck}+SH0ES best-fit early dark energy model we find a $2.6\sigma$ shift ($\Delta H_0 = 2.6\hun$) between the two measurements for Euclid; if we instead assume the ACT best-fit model, this increases to $9.0\sigma$ ($\Delta H_0 = 7.8\hun$). 
\end{abstract}

\maketitle

\section{Introduction}\label{sec: intro}
The Hubble parameter encodes the Universe's expansion rate and sets the scale of the cosmos. One of the most discussed problems in cosmology today is that two of the most precise measurements of the Hubble constant are in disagreement. In particular, constraints derived from the Cosmic Microwave Background (CMB) power spectrum give $H_0\approx 68 \hun$ \citep[e.g.,][]{2020A&A...641A...6P}, whereas local, direct, measurements of the expansion from Cepheid-calibrated supernovae give $H_0 \approx 73 \hun$ \citep[e.g.,][]{2019ApJ...876...85R,2020AJ....160...71S,Riess2021,Riess2022}. The tension between these two measurements has reached high significance ($5 \sigma$ with the most recent \textit{Planck} \cite{2020A&A...641A...6P} and SH0ES \cite{Riess2022} measurements), motivating a wealth of experimental and theoretical activity to resolve this apparent problem.

Of course, CMB and Cepheid-calibrated local measurements are not the only methods by which $H_0$ can be constrained; in particular, large-scale structure has recently emerged as a competitive probe of the expansion rate, using several methods \citep[e.g.,][]{2017MNRAS.470.2617A,2019JCAP...10..044C,2018MNRAS.480.3879A,2019ApJ...874....4A,2019JCAP...10..029S,2018ApJ...853..119A,2020JCAP...06..001C,2020arXiv200708991E,2020JCAP...05..005D,2020JCAP...05..032P,2021arXiv211007539Z,chen21}. All the measurements that are indirect (relying on a cosmological model depending on physics from both low and high redshifts) give $H_0\approx 68 \hun$. The direct measurements are somewhat less consistent: those calibrated from stellar modeling (i.e.\ tip of the red giant branch stars) fall somewhat lower than the Cepheid results \citep{Riess2021,2019ApJ...882...34F}, and, while strong lensing had previously favored a higher value, a detailed treatment of uncertainties in the mass density profiles has made this less certain \citep{2020arXiv200702941B}. Until recently, indirect measurements all had one thing in common: they derived their $H_0$ constraints from measurements of the angular scale (or redshift separation) of the sound horizon scale $r_s$ at photon-baryon decoupling ($z\approx 1100$), which was assumed to be computable via standard $\Lambda$CDM physics.\footnote{For the purposes of this work, there is little difference between `recombination' and `decoupling'; we thus use the terms interchangeably.}

This observation has led to some of the best-motivated theoretical attempts to resolve the so-called Hubble tension: invoking a mechanism that changes the sound horizon scale, $r_s$, in the early Universe. However, several hundred such models have now been proposed \cite{Knox2020,DiValentino2021}; arguably, none are perfectly well motivated, natural, and able to consistently fit the wealth of cosmological data available \cite{2020PhRvD.102d3507H,Hill2021}. Whilst one possible approach is to try and constrain each proposed model individually, we believe that this zoo of possible theories motivates the development of general diagnostics that are sensitive to such $r_s$-varying models.

Our previous work \cite{Baxter2020,Philcox2020} proposed one such consistency test: a method to measure the Hubble constant that does not rely on $r_s$, but instead makes use of another scale imprinted in the large-scale structure, namely the matter-radiation equality scale, $k_{\rm eq}$, corresponding to the size of the horizon when the density of matter and radiation were equal, at $z\approx 3600$. This scale can be given approximately within the $\Lambda$CDM model as 
\beq
    k_\mathrm{eq} = \left(2\Omega_{m}H_0^2z_\mathrm{eq}\right)^{1/2},\,\, z_\mathrm{eq} = 2.5\times 10^{4}\Omega_{m} h^2\Theta_{2.7}^{-4}
\eeq
\citep{1998ApJ...496..605E,2019JCAP...11..034C} so that $k_{\rm{eq}} \propto \omega_m h^{-1}$ when measured in units of $\hMpc$. Most directly, $k_{eq}$ sets the scale of the matter and galaxy power spectrum peak, but it also is encoded in the full shape of the spectrum around the peak (as well as, logarithmically, in the spectrum shape at higher $k$). Therefore, even though resolving the peak itself is generally difficult in galaxy surveys, $k_\mathrm{eq}$ can still be inferred from somewhat smaller \resub{(though still linear)} scales as discussed in Ref.~\cite{Philcox2020}. If observational systematic uncertainties arising, for example, from spatially varying sample completeness can be controlled on the relevant scales, $k_\mathrm{eq}$ can hence be constrained from the galaxy power spectrum.

Previously, our method avoided making use of information from $r_s$ by carefully choosing which sources of information to omit; in particular, by ensuring that our LSS analyses did not use a prior on the baryon density (unlike standard approaches), we showed that the sound horizon remained uncalibrated and hence uninformative. Of course, omitting sources of information invariably degrades the achievable constraints, which was an unfortunate feature of our method.

In this paper, we develop an analysis technique that avoids unnecessary degradation of our $H_0$ constraints by directly marginalising over templates that capture the power spectrum features encoding the sound horizon scale, namely oscillatory baryon acoustic oscillation (BAO) features and a broadband baryonic suppression. Using this framework, we are free to add additional sources of information, such as priors on the baryon density, which helps to maximise the constraining power of our new `standardisable ruler', $k_{\rm eq}$.

We demonstrate our method in a mock analysis of the Euclid spectroscopic data, forecasting the expected constraints, and also apply it to current BOSS data. Furthermore, we discuss the power of this method to constrain new physics models, using the example of early dark energy (EDE) \cite{Poulin2019,Smith2020,Lin2019,Lin2020,Sakstein2020,Niedermann2021,Hill2021}. This was recently advocated as a possible resolution to the so-called Hubble tension, though this is disputed when galaxy survey data is also included \cite{2020PhRvD.102d3507H,Ivanov2021ede,DAmico2021,Niedermann2021b,Smith2021}.

Our paper is organised as follows. In \S\ref{sec: method} we introduce our method for marginalising over the sound horizon, which is tested in \S\ref{sec:test_forecast}, using a mock-based Euclid forecast. \S\ref{sec: new-models} discusses how our new prescription can be used to test for new physics models. \S\ref{sec: boss} presents constraints with current BOSS galaxy data, before we conclude in \S\ref{sec: conclusion}. \resub{Appendices \ref{app:rs_braodband_suppression}-\ref{app:kmax_z_vary} give supplementary details relating to our procedure to marginalise over broadband information, `de-wiggled' constraint plots, and discussion of the dependence of our results on wavenumber and redshift cuts.}


\section{Measuring $H_0$ via $r_s$-marginalisation}\label{sec: method}

As previously discussed, the most promising approaches for cosmological resolutions to the so-called Hubble tension involve changing the physical size of the sound horizon, $r_s$, at photon-baryon decoupling. The large-scale distribution of matter exhibits two features controlled by $r_s$. Primarily, $r_s$ controls the BAO scale observable from the large scale structure (LSS) power spectrum through the spacing of the characteristic oscillatory feature. Additionally, $r_s$ controls the broadband baryon suppression scale, which suppresses the power spectrum on scales smaller than the sound horizon.

Given knowledge of the physical size of the sound horizon, one can use the observed angular scale of the BAO feature to constrain the Universe's expansion history. As discussed in \S\ref{sec: intro}, our desire in this work is to avoid using this information to constrain $H_0$, instead making use of the second scale available from the power spectrum; the horizon size at matter-radiation equality, $k_{\rm{eq}}$. The sound horizon, $r_s$, is not a direct input to the theory computation but rather an emergent quantity determined within a given cosmological model by integrating the sound speed, $c_s$, in the photon-baryon plasma up to decoupling
\beq
r_s = \int_0^{t_d} \frac{c_s(t)}{a(t)}dt = \int_{z_d}^\infty \frac{c_s(z)}{H(z)}dz.
\eeq
Hence, one cannot directly marginalise over it. However, we show here that such a marginalisation can be performed heuristically, by scaling the spacing of the BAO wiggles.\footnote{We return to the issue of broadband suppression below.} While such a rescaling is not physical (i.e.\ the rescaled spectra do not satisfy the perturbation equations), this is allowable: our marginalisation is a conservative approach that effectively integrates over any phenomenon capable of rescaling the BAO feature, whether or not it is allowed by physical arguments.

To implement this, we first split the power spectrum into `wiggly' and `non-wiggly' pieces, denoted $P_{\rm{nw}}(k)$ and  $P_{\rm{w}}(k)$ respectively. Such a decomposition is already necessary for the process of infrared resummation (the treatment of long wavelength displacements needed for accurate perturbative modeling of the the BAO feature \cite{ivanov_ir18,Chudaykin2020}). This separates out the physical contributions: information about the BAO wiggle spacing (and thus $r_s$) appears in the `wiggly' part, while the `non-wiggly' component contains information from the equality scale. We can thus effectively marginalise over the sound horizon by introducing a scaling parameter to the `wiggly' component only, making the replacement
\beq
P^{\rm{lin}}(k) \to P^{\rm{lin}}(k,\alpha_{r_s})\equiv P_{\rm{nw}}^{\rm{lin}}(k) + P_{\rm{w}}^{\rm{lin}}(\alpha_{r_s}k). \label{eq:rescaling}
\eeq
Notably, we perform this operation on the linear power spectrum. This is in contrast to traditional BAO analyses, which rescale the nonlinear power spectrum; we use this approach since we are interested in marginalising over a change in the physical size of the sound horizon rather than an Alcock-Paczynski scaling from cosmological coordinate conversion.

The convolution integrals relevant for the postlinear order terms in the power spectrum model (cf.\,\S\ref{subsec: pk-model}) are then evaluated using this modified linear power spectrum, $P^{\rm lin}(k,\alpha_{r_s})$. This is performed using \texttt{CLASS-PT}, a modified version of the Boltzmann code \texttt{CLASS} that has has been augmented with a perturbation theory treatment of nonlinear structure formation (see Ref.~\cite{Chudaykin2020} for details). We have further extended this to implement the above $r_s$-marginalisation procedure.\footnote{\texttt{CLASS-PT} is available online here: \href{https://github.com/Michalychforever/CLASS-PT}{GitHub.com/Michalychforever/CLASS-PT} and our modified version may be accessed at \href{https://github.com/gerrfarr/CLASS-PT}{GitHub.com/gerrfarr/CLASS-PT}.} Within \texttt{CLASS-PT} separation of the power spectrum components is performed in the following way. First, the spectrum is transformed to position space via a discrete sine transform, the BAO feature is removed and the resulting correlation function is smoothly interpolated (cf.\,\S4 of Ref.~\cite{Chudaykin2020}). Converting back to momentum space one obtains the `non-wiggly' power spectrum. The `wiggly' part (before $\alpha_{r_s}$-rescaling) is equal to the difference between the initial linear power spectrum and the `non-wiggly' part, $P^{\rm{lin}}_{\rm{w}}(k) \equiv P_{\rm{nw}}^{\rm{lin}}(k) - P^{\rm{lin}}(k)$.

\begin{figure}[ht!]
    \centering
    \includegraphics[width=\linewidth]{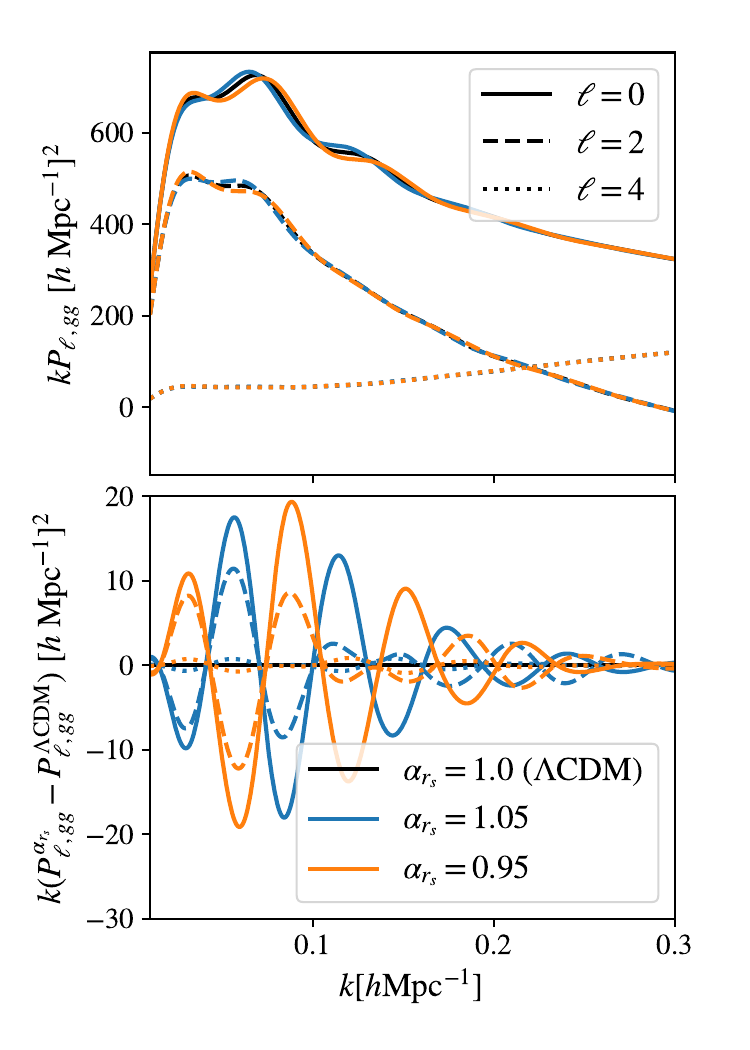}
    \caption{\textbf{Upper panel:} Monopole, quadrupole and hexadecapole of the galaxy power spectrum for different values of the sound horizon rescaling parameter $\alpha_{r_s}$. This rescales the `wiggly' component of the power spectrum only (see Eq.\,\ref{eq:rescaling}), facilitating marginalisation over the sound horizon $r_s$. We assume a Euclid-like spectroscopic sample similar to that of Ref.~\cite{Chudaykin2019} and neglect the stochastic (shot-noise) contribution to the monopole for the purpose of these plots. Here, we show the spectra in the lowest redshift bin, centered on $z=0.6$. \textbf{Lower panel:} The characteristic BAO wiggles are shifted to larger scales for $\alpha_{r_s}>1$ and to smaller scales for $\alpha_{r_s}<1$. On small scales, the modified spectra approach the $\Lambda$CDM scenario given that we have not modified the $r_s$ dependent small scale baryonic suppression in this figure.}
    \label{fig:pk_alpha_rs}
\end{figure}

Fig.\,\ref{fig:pk_alpha_rs} shows model nonlinear power spectra for different values of $\alpha_{r_s}$. One can clearly observe that the action of the rescaling parameter is to shift the characteristic BAO wiggles, as required. On scales $k\gtrsim 0.2 h\rm{Mpc}^{-1}$ the wiggly feature is strongly suppressed, thus the rescaled spectra agree with the fiducial one, as expected. In \S\ref{sec:test_forecast} we discuss our nonlinear power spectrum model, free parameters and fiducial assumptions in more detail.

The baryon suppression scale provides a second source of information about $r_s$, which is nonoscillatory, and will thus not be marginalised over by the above prescription. Instead, we adopt an approximate procedure to remove the information, using a method similar to that applied to nuisance parameter marginalisation in Refs. \cite{Chudaykin2019,om4}. To motivate this, we consider some parameter $\theta$, with a Gaussian prior of mean $\bar\theta$ and width $\sigma_\theta$, entering our model vector $\bm{m}$ linearly. This can be exactly marginalised over by evaluating $\bm{m}$ at $\bar\theta$, and modifying the data covariance as follows \cite{2010MNRAS.408..865T}:
\beq
C_{\rm{new}} = C_D + \sigma_\theta^2 \left[\dv{\bm{m}}{\theta}\right]\left[\dv{\bm{m}}{\theta}\right]^T.
\eeq

While the nonlinear power spectrum clearly does not depend linearly on a rescaling of the baryon suppression scale (denoted by $\beta_{r_s}$), we can Taylor expand our model to first order in this rescaling 
\beq
\bm{m} = \bm{m}(\beta_{r_s}=1) + \dve{\bm{m}}{\beta_{r_s}}{\beta_{r_s}=1} \left(\beta_{r_s} - 1\right) + \mathcal{O}(\beta_{r_s}^2).
\eeq
An approximate marginalisation at lowest order is thus possible analytically even for nonlinear dependencies. In principle one might envision using a similar procedure to also marginalise over the BAO feature in the power spectrum. However, in practice the linear approximation breaks down even for relatively small rescalings of the oscillatory BAO feature. The much smoother power spectrum features associated with the baryon suppression scale are more accurately approximated by a linear order expansion.\footnote{In initial testing, the BAO feature was marginalised over via this method, obtaining model derivatives using the analytic Eisenstein-Hu model \cite{1998ApJ...496..605E} as described here for the baryon suppression scale. This method was not able to achieve constraints largely independent of $r_s$ and hence we adopted the approach described in the main text.}

To obtain the relevant derivatives in practice, we consider the analytic Eisenstein-Hu transfer function \citep{1998ApJ...496..605E}. This model smoothly interpolates between an unsuppressed transfer function $T_{\rm{usp}}(k)$ and a baryon-suppressed transfer function $T_{\rm{sp}}(k)$ near the sound horizon scale using
\beq
T(k) = f(k) T_{\rm{usp}}(k) + \left[1-f(k)\right]T_{\rm{sp}}(k)
\eeq
where $f(k) = (1 + k r_s/C)^\alpha$ for empirical constants $C$ and $\alpha$. To rescale the suppression scale we make the replacement $(1 + k r_s/C)^\alpha \to (1 + \beta_{r_s} k r_s/C)^\alpha$. Given this modified linear power spectrum, we differentiate our full nonlinear model with respect to $\beta_{r_s}$ numerically, using a five-point, finite difference rule. In Appendix \ref{app:rs_braodband_suppression} we demonstrate, however, that, at least for a Euclid-like survey with our current modeling choices, the impact of the broadband-derived $r_s$ information is negligible, so that we can ignore it for the remainder of our paper.

\section{Method Verification and Euclid Forecasts} \label{sec:test_forecast}

In this section, we present the results of Markov chain Monte Carlo (MCMC) analyses, which aim to demonstrate that the method described in \S\ref{sec: method} can produce $r_s$-independent constraints on $H_0$. Furthermore, we demonstrate that for future spectroscopic surveys such as Euclid, such constraints will be competitive with those from other probes.

\subsection{Power spectrum model}\label{subsec: pk-model}

Given that one of the goals of this paper is to provide a partially model-independent test for various cosmological models that propose solutions to the so-called Hubble tension, we fix our baseline cosmology to the best fit $\Lambda$CDM model from Ref.~\cite{Smith2020}, which will allow us to later contrast the results with their best fit early dark energy (EDE) model. Explicitly, our fiducial cosmology is given by 
\begin{eqnarray*}
 H_0 &=& 68.21\hun\\
 \omega_{\rm{cdm}} &=& 0.1177\\ 
 \omega_{\rm{b}} &=& 2.253 \times 10^{-2}\\
 n_s &=& 0.9686\\
 A_s &=& 2.216 \times 10^{-9}\\
 \tau_{\rm{reio}} &=& 0.085
\end{eqnarray*}
We also include a single massive neutrino species with mass $0.06\rm{\ eV}$.

Our standard analysis assumes a Euclid-like spectroscopic survey, and models the power spectrum using the effective field theory of LSS. This features a consistent one-loop perturbative model, including ultraviolet counterterms, infrared resummation, Alcock-Paczynski distortions and Fingers-of-God corrections \citep[cf.,][]{Ivanov2017,2020JCAP...05..005D,chen21}. Our power spectrum model contains nine nuisance parameters: the linear, quadratic, tidal and cubic tidal bias parameters ($b_1$, $b_2$, $b_{\mathcal{G}_2}$ and $b_{\Gamma_3}$), the monopole, quadrupole and hexadecapole counterterms ($\tilde{c}_0$, $\tilde{c}_2$ and $\tilde{c}_4$), and two stochastic contributions, ($P_{\rm{shot}}$ and $a_2$), scaling as $k^0$ and $k^2$ respectively. For details, we refer the reader to Refs.~\cite{Chudaykin2020,Ivanov2017}, or Ref.~\cite{Chudaykin2019}, for those pertaining to Euclid. We additionally assume a Gaussian likelihood, with a theoretical covariance matrix, encoding both cosmic variance and theoretical uncertainties, following Ref.~\cite{Chudaykin2019}.

The survey is specified by eight nonoverlapping redshift bins evenly spaced between $z_{\rm{min}} = 0.5$ and $z_{\rm{max}}=2.1$. Within each redshift bin we analyse the monopole, quadrupole and hexadecapole in 40 bins evenly spaced in $\log k$ between $k_{\rm{min}}=0.01 \hMpc$ (matching past BOSS analyses) and $k_{\rm{max}}=1.0 \hMpc$. In addition to our cosmological parameters (and where relevant the sound horizon marginalisation parameter, $\alpha_{r_s}$), we include a total of 72 nuisance parameters (nine for each redshift slice). To reduce the dimensionality of our parameter space, we analytically marginalise over any parameter that enters our model linearly \citep{Chudaykin2019,om4}; these are the counterterm parameters, the stochasticity parameters, and the cubic tidal bias parameters. This procedure is exact and equivalent to numerical marginalisation with Gaussian priors \cite{2010MNRAS.408..865T}. The remaining bias parameters, $b_1$, $b_2$ and $b_{G_2}$, are allowed to vary freely, though we impose Gaussian priors of width $\sigma_{b_X}=1$ on the latter two, centred on the true values.

Mock power spectrum data are generated using the following fiducial values for the nuisance parameters:
\begin{widetext}
\beq
\begin{aligned}[c]
b_1 &= 0.9 + 0.4 z\\
b_{\mathcal{G}_2} &= \frac{2}{7}\left(1-b_1(z)\right)\\
b_2 &= -0.704 - 0.208 z + 0.183 z^2 - 0.0077 z^3 + \frac{4}{3}b_{\mathcal{G}_2}(z)\\
b_{\Gamma_3} &= \frac{23}{42}\left(b_1(z) -1\right)\\
\end{aligned}
\qquad
\begin{aligned}[c]
\tilde{c}_0 &= 1.9 D_+^2(z) [\Mpch]^2\\
\tilde{c}_2 &= 52 D_+^2(z) [\Mpch]^2\\
\tilde{c}_4 &= -2.4 D_+^2(z) [\Mpch]^2\\
a_2 &= 0\\
P_{\rm{shot}} &= n_g(z)^{-1},
\end{aligned}
\eeq
\end{widetext}
where $n_g(z)$ is an estimate of the number density of galaxies observed in each redshift bin and $D_+(z)$ is the scale independent growth rate computed with our fiducial cosmology. Note that the values adopted for the counterterm parameters are based on fits to the eBOSS ELG sample \cite{Ivanov2021}, rather than those presented in Ref.~\cite{Chudaykin2019}. We do not expect the fiducial choice for these nuisance parameters to significantly affect our forecasts. The fiducial numeric values for all nuisance parameters are shown in Table\,\ref{tab:nuisance_params}. Furthermore, we note that no BAO reconstruction is included in this analysis; this would further tighten the $r_s$-based constraints. We do not include noise in our mock data. 

\begin{table*}
\caption{Fiducial numerical values for the 64 nonzero nuisance parameters, as well as the galaxy number density (in $h^3 \rm{Mpc}^{-3}$ units) and bin volume (in $h^{-3} \rm{Gpc}^3$ units). The scale-independent stochastic contribution is shown in $h^{-3} \rm{Mpc}^{3}$ units and the counterterm parameters in units of $h^{-2}\rm{Mpc}^2$. The $a_2$ parameter has a fiducial value of $0$, and is not shown. Note that the number densities and bin volumes do not agree exactly with those from Ref.~\cite{Chudaykin2019}; this is a consequence of slightly different fiducial cosmology. Furthermore, we adopt the counterterm parameters fit to the eBOSS ELG sample, as in Ref.~\cite{Ivanov2021}. We marginalise over all 72 nuisance parameters in our analysis, employing analytic marginalisation with conservative priors over the counterterm parameters, the stochasticity parameters, and the cubic tidal bias parameters.\label{tab:nuisance_params}}
\centering
\begin{tabular}{d | d | d | d d d d | d d d | r d }
\hline\hline
\multicolumn{1}{c|}{\quad $\bar{z}$ \quad } & \multicolumn{1}{c|}{\quad $10^3 n_g(\bar{z})$\quad }& \multicolumn{1}{c|}{\quad $V(\bar{z})$\quad }&
\multicolumn{1}{c}{\quad $b_1$\quad } & \multicolumn{1}{c}{\quad $b_2$\quad } & \multicolumn{1}{c}{\quad $b_{G_2}$\quad } & \multicolumn{1}{c|}{\quad $b_{\Gamma_3}$\quad } & \multicolumn{1}{c}{\quad $\tilde{c}_0$\quad }& \multicolumn{1}{c}{\quad $\tilde{c}_2$\quad } & \multicolumn{1}{c|}{\quad $\tilde{c}_4$\quad } & \multicolumn{1}{c }{\quad $P_{\rm{shot}}$\quad } & \multicolumn{1}{c }{\quad $k_{\rm{shot\ dom.}}$\quad }  \\ \hline
0.6 & 3.75 & 4.68  & 1.14 & -0.82 & -0.04 & 0.08 & 1.03 & 28.15 & -1.30 & 267 & 0.49   \\
0.8 & 2.03 & 6.61  & 1.22 & -0.84 & -0.06 & 0.12 & 0.85 & 23.28 & -1.07 & 493 & 0.43  \\
1.0   & 1.15 & 8.25  & 1.30 & -0.85 & -0.09 & 0.16 & 0.71 & 19.46 & -0.90 & 872 & 0.32  \\
1.2 & 0.67 & 9.55  & 1.38 & -0.85 & -0.11 & 0.21 & 0.60 & 16.45 & -0.76 & 1484 & 0.22 \\
1.4 & 0.38 & 10.53 & 1.46 & -0.83 & -0.13 & 0.25 & 0.51 & 14.05 & -0.65 & 2643 & 0.15 \\
1.6 & 0.20 & 11.23 & 1.54 & -0.81 & -0.15 & 0.30 & 0.44 & 12.11 & -0.56 & 4943 & 0.10 \\
1.8 & 0.11 & 11.71 & 1.62 & -0.77 & -0.18 & 0.34 & 0.39 & 10.54 & -0.49 & 8865 & 0.06 \\
2.0   & 0.07 & 12.02 & 1.70 & -0.72 & -0.20 & 0.38 & 0.34 & 9.25  & -0.43 & 15323& 0.03\\\hline\hline
\end{tabular}
\end{table*}

\subsection{Parameter recovery and forecasts}\label{subsec: validation}

Reference~\cite{Philcox2020} noted that $r_s$-independent constraints can be derived from a full shape (FS) galaxy power spectrum analysis when calibration of the BAO feature is explicitly avoided, by removing prior information on the baryon density, $\omega_{\rm b}$. In our analyses, we can either omit this prior as before or include it while marginalising over $r_s$; we present results from a full MCMC analysis for both cases.

\begin{table}[]
\caption{$H_0$ constraints (in $\hun$ units) from the Euclid spectroscopic forecast, assuming an underlying $\Lambda$CDM cosmology based on \citep{Smith2020}. Results are shown for three choices of likelihood: (1), the full-shape likelihood (FS), which captures all power spectrum information, (2), the full-shape likelihood, marginalised over $r_s$ using the method of \S\ref{sec: method} (FS + $r_s$ marg.), and (3), an $r_s$-only likelihood (BAO). These source $H_0$ information from $k_{\rm eq}$ and $r_s$, $k_{\rm eq}$, and $r_s$ respectively. We consider analyses with and without a BBN-based prior on $\omega_{\rm b}$. In each case, we report the $68\%$ confidence interval on $H_0$ in units of $\hun$. The `FS + BBN + $r_s$-marg.' constraint -- which is independent of $r_s$ -- is the main result of this work.}\label{tab:h0-posteriors}
    \centering
    \begin{tabular}{l|c}
    \hline\hline
     & $H_0$ \\\hline
     FS & $68.1^{+1.2}_{-1.6}$\\
     FS + $r_s$ marg. & $68.0^{+1.5}_{-2.0}$ \\
     FS + BBN & $68.17\pm 0.40$ \\
     FS + BBN + $r_s$ marg. & $\mathbf{68.15\pm 0.72}$\\
     BAO + BBN & $68.28\pm 0.49$ \\
     BAO + BBN + $r_s$ marg & $68.8^{+1.4}_{-1.6}$\\
     \hline\hline
    \end{tabular}
\end{table}


\begin{figure}[ht!]
  \centering
    \includegraphics[width=\linewidth]{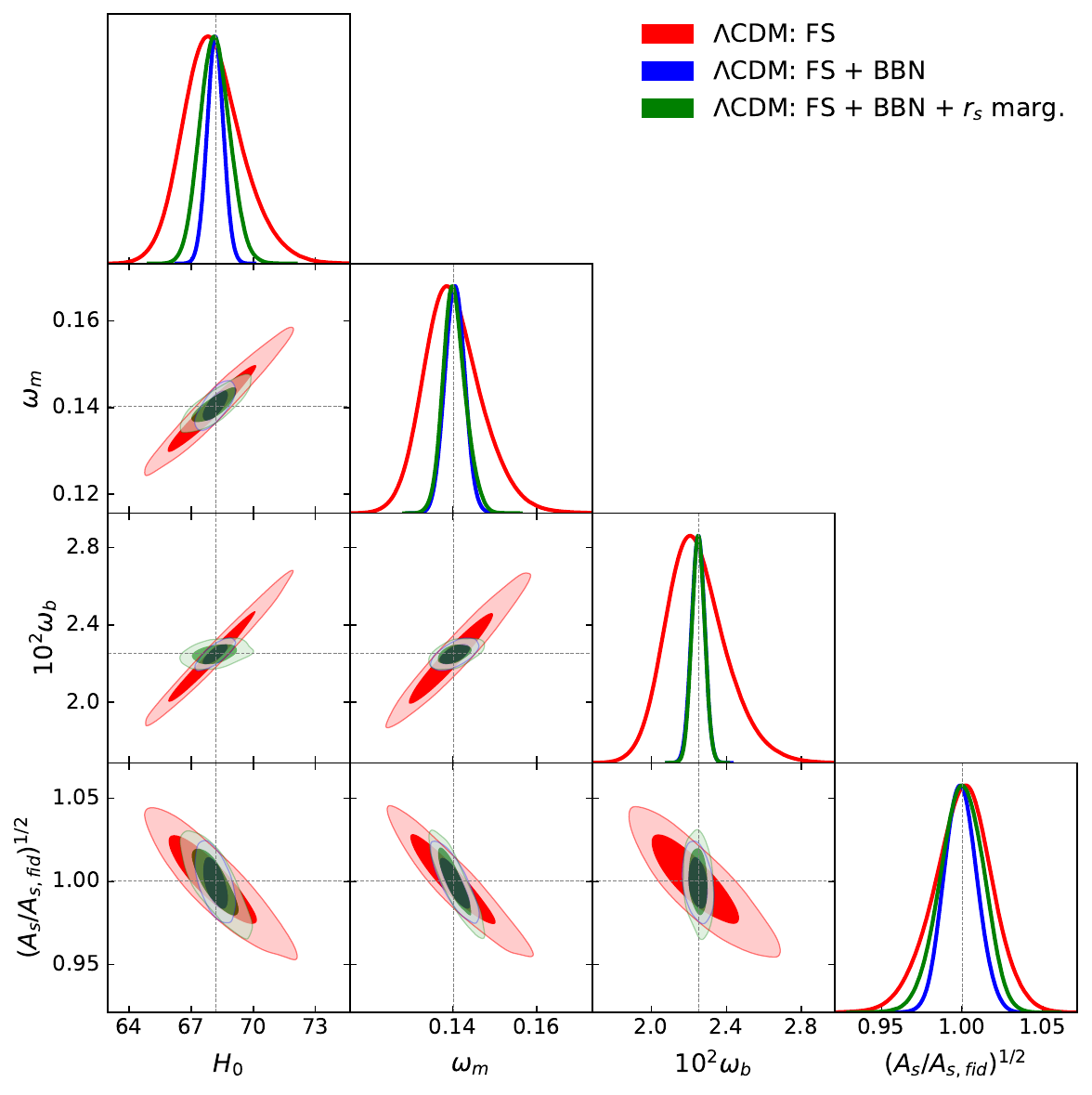}
    \caption{Forecasted parameter constraints from a power spectrum analysis of a Euclid-like survey, assuming a $\Lambda$CDM cosmology. Three datasets are shown: the full power spectrum likelihood (red), the full power spectrum likelihood with the addition of a BBN prior on $\omega_{\rm b}$ (blue), and the same with additional marginalisation over the sound horizon $r_s$ (green, see \S\ref{sec: method}). The resulting constraints on $H_0$ are given in Table\,\ref{tab:h0-posteriors}. The third quantity (in green) is the main result of this work: an $r_s$-independent constraint on $H_0$ that excludes information from the sound horizon. This is significantly narrower than that from the first dataset (red), which represents the previous $r_s$-independent approach proposed in Ref.~\citep{Philcox2020}.}
    \label{fig:lcdm_oliver_plot}
\end{figure}

\begin{figure}
  \centering
  \includegraphics[width=\linewidth]{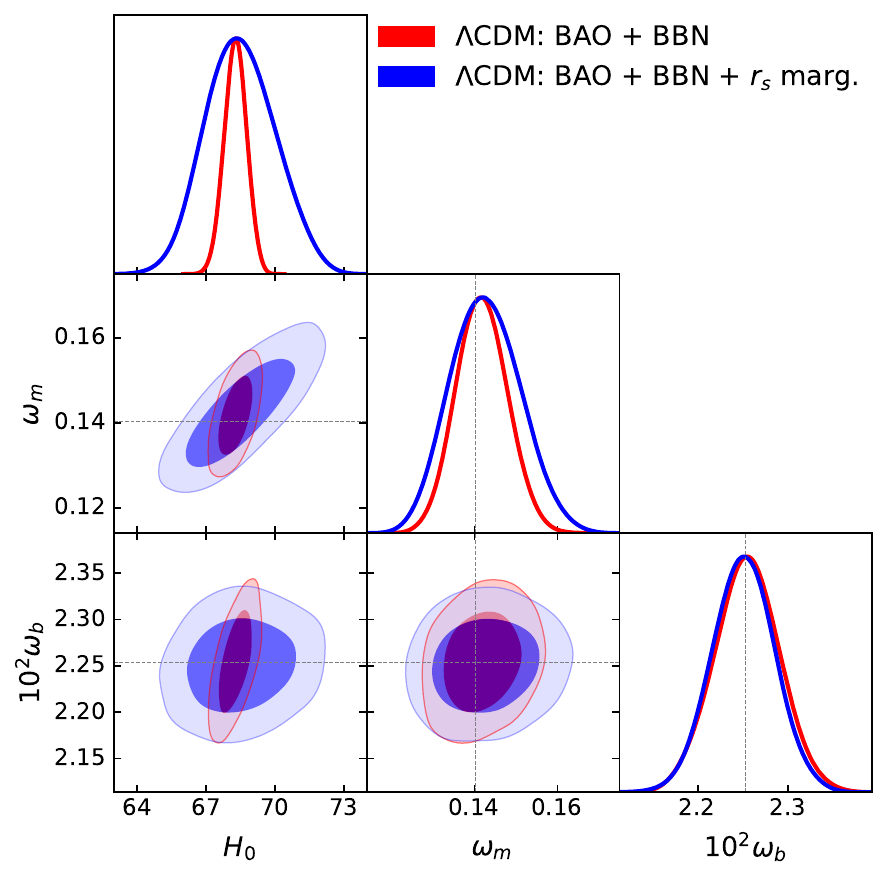}
  \caption{Parameter constraints from a BAO-only analysis of the Euclid mock data, generated using a $\Lambda$CDM cosmology. The action of $r_s$-marginalisation (\S\ref{sec: method}) inflates the $H_0$ posterior by about a factor of 3, from $68.28\pm 0.49$ to $68.5^{+1.4}_{-1.6}$ (in $\hun$ units). The large width of this parameter indicates that the marginalisation is working successfully -- while the error is not infinite, the marginalisation performs well enough that the sound horizon is a negligible source of $H_0$ information compared with the equality scale. Note that the BAO-only likelihood cannot constrain $A_s$ thus no constraints are shown for the parameter.}
  \label{fig:lcdm_BAO}
 \end{figure}

\begin{figure*}[t!]
    \subfloat[MCMC forecasts]{%
      \includegraphics[width=0.48\textwidth]{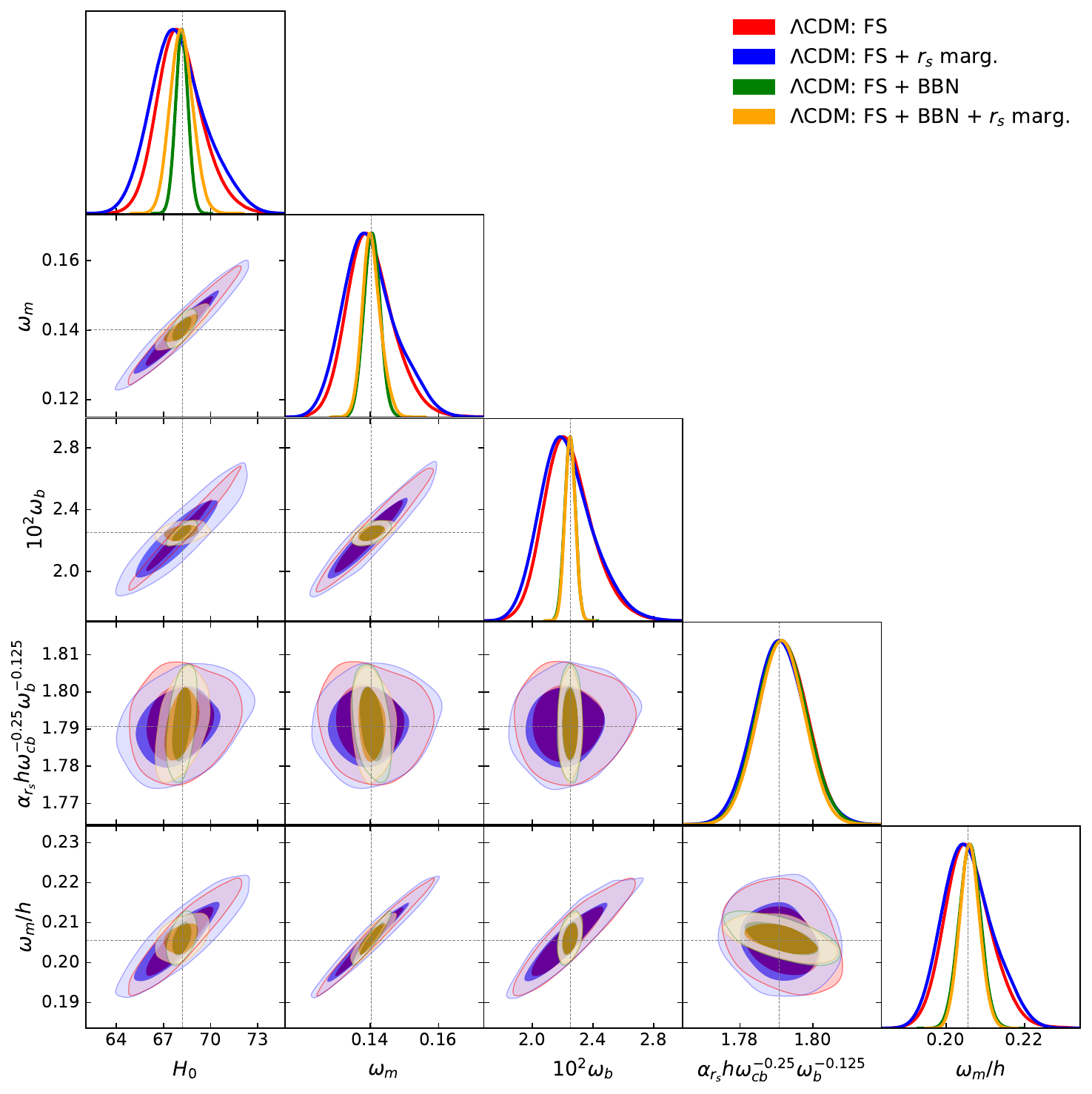}\label{subfig:MCMC_forecasts}
    }
    \hfill
    \subfloat[Fisher forecasts]{%
      \includegraphics[width=0.48\textwidth]{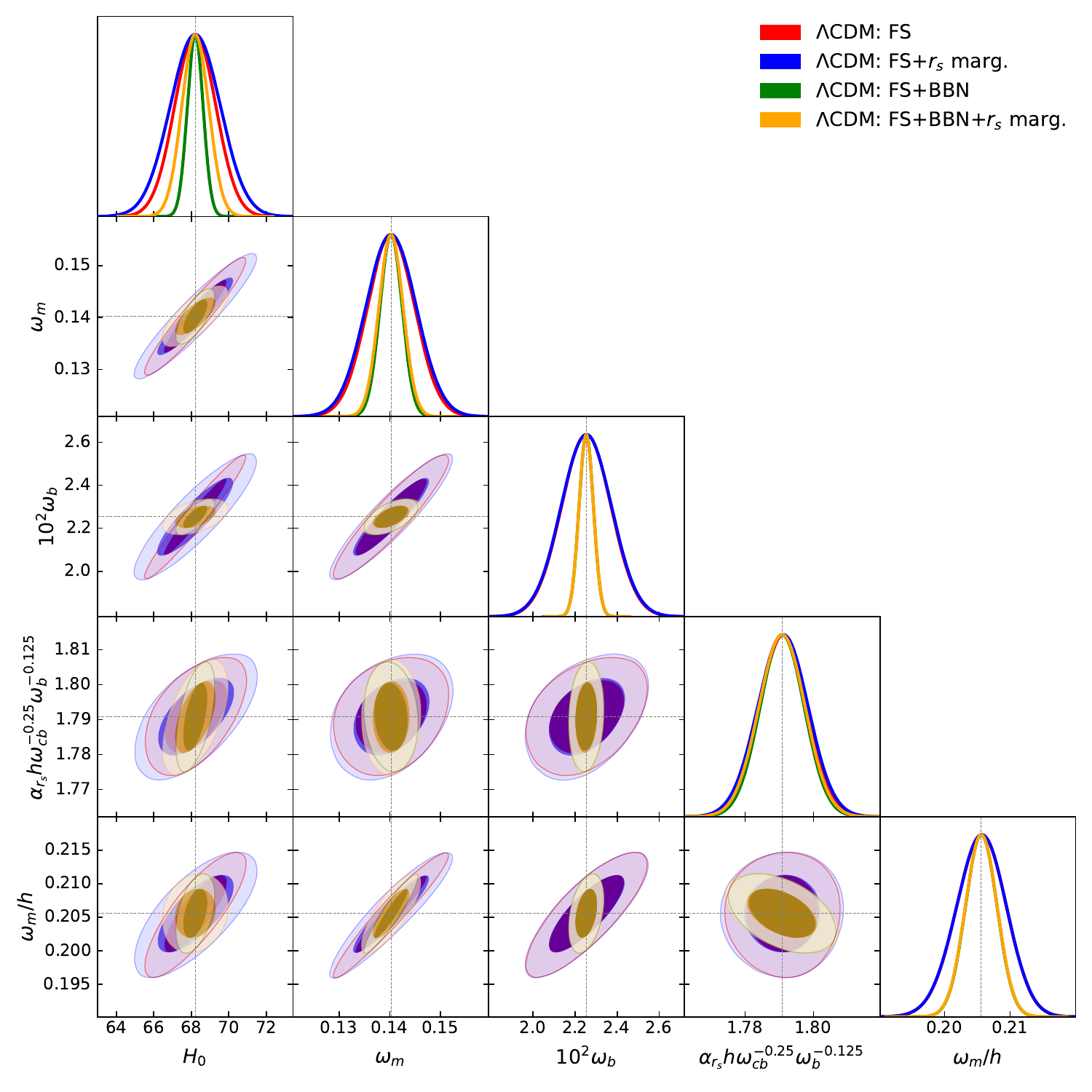}\label{subfig:fisher_forecasts}
    }
    \caption{Comparison of parameter posteriors from a full MCMC forecast using the heuristic marginalisation procedure described in \S\,\ref{sec: method} (left) and a Fisher forecasts using an Eisenstein-Hu \cite{1998ApJ...496..605E} model with exact marginalisation over the sound horizon scale (right), including the suppression of power on scales smaller than $r_s$. We find excellent agreement between the two posteriors, both in terms of the degeneracy directions and parameter constraints, suggesting that our approach is working as expected. This is further supported by consideration of the degeneracy direction between $H_0$ and the $r_s$ proxy \eqref{eq: rs-proxy}, as discussed in \S\ref{subsec: rs-tests}.}
    \label{fig:fisher_comparison}
\end{figure*}

Results from the various MCMC analyses are shown in Fig.\,\ref{fig:lcdm_oliver_plot} and Table\,\ref{tab:h0-posteriors}. First, we consider the results obtained using the previous prescription: when performing an analysis without a baryon prior or $r_s$-marginalisation, we obtain a $\sim 2\%$ measurement of $H_0$ ($H_0 = 68.1^{+1.2}_{-1.6} \hun$), in good agreement with our earlier forecast. The omission of such a prior, however, not only removes information from the BAO scale but also reduces the information that can be extracted from other features of the FS power spectrum. Figure\,\ref{subfig:MCMC_forecasts} shows a clear degeneracy between $H_0$ and $\omega_b$ even for the $r_s$-marginalised analysis. This is not unexpected; clearly any additional information on $\omega_b$ will directly improve constraint on $\omega_m=\omega_{\rm{cdm}} +\omega_{b} + \omega_\nu$ (and thus $k_{\rm eq}\propto \omega_m h^{-1}$ in $\hMpc$ units). Additionally, the amplitudes of the BAO peaks, for example, are sensitive to the ratio of baryons to dark matter and can thus be used to sharpen the constraint on $\omega_m$ further if information about the baryon density is provided.

When a prior on the baryon density is included [specifically, a Gaussian prior of $\omega_{\rm b} = (2.253 \pm 0.036) \times 10^{-2}$ from big bang nucleosynthesis; BBN],\footnote{This has the same fractional width as in Ref.~\cite{2020JCAP...05..032P}.} our forecasted constraint tightens to $H_0 = 68.17\pm 0.40 \hun$. However, this will now include information sourced by $r_s$. Using the $r_s$-marginalisation procedure described in \S\ref{sec: method} (integrating over $\alpha_{r_s}$), we can isolate the $k_{\rm eq}$-based information (ignoring broadband suppression, which Appendix \ref{app:rs_braodband_suppression} finds to be negligible), which yields $H_0 = 68.15 \pm 0.72\hun$. This is substantially tighter than our analysis without BBN information, and shows the utility of our method. In \S\ref{subsec: rs-tests} we demonstrate that this is indeed independent of $r_s$. It should be noted that our constraints without the BBN prior are somewhat wider than those reported in Ref.~\cite{Chudaykin2019}. This occurs since our work is based on an updated version of the relevant Euclid likelihood, which, for example, adds the scale-dependent stochastic contribution $a_2$. Furthermore, we can quantify the information content of $r_s$ alone with a BAO-only forecast, akin to that done in traditional power spectrum analyses \citep[e.g.][]{2017MNRAS.470.2617A}. This is described in \S\ref{subsec: rs-tests}, and gives the constraint $H_0 = 68.28\pm0.49 \hun$, somewhat stronger than the equality-based constraint. Notably, the combined constraint is approximately equal to the inverse-variance weighted mean of the two constraints, hinting at their independence. The fact that the BAO constraint is significantly tighter than the equality-derived constraint illustrates that the combined constraint is dominated by $r_s$-derived information, with the $k_{eq}$-derived result obscured in the combination; $r_s$-marginalisation is therefore necessary to isolate the equality-derived information.

As seen in Fig. \ref{fig:lcdm_oliver_plot}, we obtain unbiased and consistent parameter recovery for all cosmological parameters both with and without a prior on $\omega_{\rm b}$. When applying our heuristic $r_s$-marginalisation procedure without the $\omega_{\rm b}$ prior we find only a very minor difference to the standard FS analysis ($H_0=68.0^{+1.5}_{-2.0}\hun$, $\Delta \sigma_{H_0} \simeq 0.4$). This is expected given that sound horizon information should be subdominant in this analysis from the outset (since the standard ruler is uncalibrated).

\subsection{$r_s$ independence}\label{subsec: rs-tests}

We now present a set of tests to validate the independence of our $H_0$ constraints from the sound horizon scale. This is necessary to ensure that we are indeed obtaining information from $k_{\rm eq}$ and that our constraints are not derived from residual $r_s$-derived information. Three different tests are presented. 

First, we compare to results from a BAO-only analysis, which does not include information from the broadband shape of the power spectrum. Next, we compare to Fisher forecasts including an exact marginalisation over the sound horizon, and, finally, inspect the degeneracy between the approximate sound horizon scale and $H_0$. In Appendix~\ref{app:no_wiggle} we also compare our analysis to one run on a dataset from which the BAO feature has been removed. We find a similar $H_0$ posterior, showing that such constraints can be derived from the broadband alone. 

\resub{Furthermore, we explore the impact of scale and redshift selections in Appendix \ref{app:kmax_z_vary} using Fisher forecasts. For this purpose we reanalyse the mock data using different $k_{\rm{max}}$ values of $0.1$, $0.5$ and $1.0 \hMpc$ and divide our sample up into low-, medium- and high-redshift datasets ($z<1.0$, $1.0\leq z<1.4$ and $1.4\leq z$ respectively). $r_s$ marginalisation is found to affect our results even for $k_{\rm{max}}=0.1\hMpc$. Although this may appear counterintuitive, since there should be little $r_s$ information below $k=0.1\hMpc$, it is expected. The Fourier-space equality peak partially overlaps with the first BAO peak; allowing the BAO scale to shift freely also degrades the equality-derived constraints by increasing the uncertainty with which the turnover scale can be measured. Additionally, excluding higher $k$ modes suppresses other information contained in the power spectrum on those scales, which constrains $\omega_m$, for example. This therefore leads to a significant degradation of $H_0$ constraints. We also find that including our $r_s$-marginalisation increases the relative weight of the the low-redshift data, for which constraints are less significantly degraded by the removal of BAO information compared to the medium- and high-redshift bins.}


\subsubsection{BAO-only analysis}
First, we run a BAO-only analysis on the same (unreconstructed) mock datasets as our FS analysis (see Fig.\,\ref{fig:lcdm_BAO}). Similarly to Ref.~\cite{2020JCAP...05..032P}, we employ a theoretical error model to effectively marginalise over the broadband power spectrum  (as proposed in Ref.~\cite{Baldauf2016}). This removes most information not present in the wiggle positions, and is practically implemented by introducing a large covariance with correlation length larger than the BAO scale. To perform the analysis, we vary $\omega_{\rm b}$, $\omega_{\text{cdm}}$ and $h$, as well as the same nuisance parameters as above. Utilising the previous BBN-derived prior on $\omega_{\rm b}$ we find $H_0 = 68.28 \pm 0.49 \hun$ from BAO, which is degraded by a factor close to three (to $H_0=68.5 ^{+1.4}_{-1.6} \hun$), when the sound horizon rescaling parameter $\alpha_{r_s}$ is also marginalised over.

Ideally, the $r_s$-marginalised BAO constraint should be infinitely wide, since we are integrating over the feature of interest. The residual, albeit weak, constraints could potentially be explained by a small amount of broadband information remaining in the analysis. To test whether the constraints had residual dependence on the broadband shape, we included a smooth polynomial in the power spectrum model (as in Ref.~\cite{Beutler2017}), whose coefficients were analytically marginalised over. Furthermore, marginalisation over an overall rescaling of the three multipoles separately in the eight redshift bins was included. These tests somewhat inflated the marginalised constraints (giving $H_0= 68.8 \pm 1.8\hun$) indicating that, indeed, some small amount of broadband information was leaking into the BAO analysis. Additionally, we note that the coordinate rescaling probed by the BAO analysis, while largely degenerate with a change in the physical size of the sound horizon, is not exactly identical to our $r_s$-marginalization operation on the linear power spectrum. Hence, some very weak residual constraints on $H_0$ might be expected. Given that the $H_0$ constraints from marginalised BAO are much wider than those from the marginalised FS pipeline, we take this as an indication that our prescription gives constraints that are effectively independent of the sound horizon -- any residual $r_s$-derived information is highly subdominant. 


\subsubsection{Comparison with forecasts including exact $r_s$-marginalisation}

In Fig.\,\ref{fig:fisher_comparison} we compare our MCMC forecasts for Euclid with corresponding Fisher forecasts using an Eisenstein-Hu model \cite{1998ApJ...496..605E} for the power spectrum, but with the same experimental setup as described in \S\ref{sec: method}. Within this model we are able to perform the marginalisation over the sound horizon exactly by modifying $r_s$ (an exact parameter within this model) within the transfer function computation \citep[cf.][Eq.\,6]{1998ApJ...496..605E}. This marginalisation also includes the effects of baryon-induced broadband suppression. Clearly our Fisher forecasts do not capture the posteriors' non-Gaussianity (visible particularly in the $H_0 - \alpha_{r_s}h \omega_{cb}^{-0.25}\omega_{\rm b}^{-0.125}$ panel). Nevertheless, the posteriors and degeneracy directions are found to be in excellent overall agreement. This is particularly true when BBN information is included, which leads to the MCMC forecasts becoming more Gaussian. From the Fisher forecasts we find a constraint on the Hubble parameter of $\sigma_{H_0}=0.72 \hun$ for our `FS+BBN+$r_s$ marg.' analysis (compared to $\sigma_{H_0}=0.43 \hun$ for the `FS+BBN' analysis) in good agreement with our MCMC results ($\sigma_{H_0}=0.72 \hun$ and $\sigma_{H_0}=0.40 \hun$ for `FS+BBN+$r_s$ marg.' and `FS+BBN' respectively). The striking agreement between the two further supports our claim of $r_s$-independence, given that the Fisher results feature exact sound horizon marginalisation.

\subsubsection{Degeneracy between $H_0$ and $r_s$}
Within $\Lambda$CDM, the sound horizon scale can be approximately written as
\beq\label{eq: rs-proxy}
r_s \simeq 55.15 h \omega_{cb}^{-0.25}\omega_{\rm b}^{-0.125} \Mpch
\eeq
\citep{2015PhRvD..92l3516A}. In our marginalised analysis, the effective sound horizon being fit thus scales as $\alpha_{r_s} h \omega_{cb}^{-0.25}\omega_{\rm b}^{-0.125}$. In Fig.\,\ref{fig:fisher_comparison} we show the degeneracy of this parameter combination with $H_0$ for the case of the `FS+BBN+$r_s$ marg.' analysis (blue contour in the leftmost panel on the second to last row in the left panel). Notably, this effective sound horizon exhibits no significant degeneracy with $H_0$, affording us confidence that the constraint obtained is in fact almost entirely derived from $k_{\rm{eq}}$ and independent of $r_s$.


\section{A Null Test for New Physics Models}\label{sec: new-models}

\begin{table}[]
\caption{$H_0$ constraints (in $\hun$ units) from the Euclid spectroscopic forecast, as in Table\,\ref{tab:h0-posteriors}, but now assuming underlying EDE cosmologies based on \citep{Smith2020} (\textit{Planck}+SH0ES) and \citep{Hill2021} (ACT). In all cases, data are analysed assuming $\Lambda$CDM. Again, results are shown for three choices of likelihood: (1) the full-shape likelihood (FS), which captures all power spectrum information; (2), the full-shape likelihood, marginalised over $r_s$ using the method of \S\ref{sec: method} (FS + $r_s$ marg.); and (3), an $r_s$-only likelihood (BAO). These source $H_0$ information from $k_{\rm eq}$ and $r_s$, $k_{\rm eq}$, and $r_s$ respectively. We consider analyses with and without a BBN-based prior on $\omega_{\rm b}$. In each case, we report the $68\%$ confidence interval on $H_0$ in units of $\hun$. We highlight in bold one of the main results of this work, the shifts between $k_{\rm eq}$- and $r_s$-based measurements of $H_0$ for mock data generated within an EDE cosmology.}\label{tab:h0-posteriors-ede}
    \centering
    \begin{tabular}{l|c|c}
    \hline\hline
    & EDE (\textit{Planck}+SH0ES) & EDE (ACT) \\\hline
     FS & $67.58^{+0.95}_{-1.4}$ & $67.08^{+0.95}_{-1.2}$\\
     FS + $r_s$ marg.  & $66.7\pm 1.8$ &$64.53^{+0.88}_{-0.74}$\\
     FS + BBN & $69.54\pm 0.45$& $73.43\pm 0.51$\\
     FS + BBN + $r_s$ marg. & $\mathbf{67.39^{+0.89}_{-0.79}}$&$\mathbf{66.82\pm 0.53}$\\
     BAO + BBN & $\mathbf{69.97\pm 0.50}$& $\mathbf{74.62\pm0.69}$\\\hline\hline
    \end{tabular}
\end{table}

Using the above techniques, we can obtain $r_s$-independent constraints on $H_0$ using the equality scale, $k_{\rm eq}$. Assuming our $\Lambda$CDM model of the Universe to be accurate, this estimate should be statistically consistent with that derived from the BAO feature; a corollary is that any \textit{difference} between the two estimates gives evidence for non-standard physics operating in the early Universe. This was originally pointed out in Ref.~\citep{Philcox2020}; in this work, we provide a practical example in the context of recently proposed physical models \citep[e.g.,][]{Smith2020}, again forecasting for the Euclid satellite.

\begin{figure*}[!p]
    \subfloat[$\Lambda$CDM: Without BBN Priors]{%
      \includegraphics[width=0.48\textwidth]{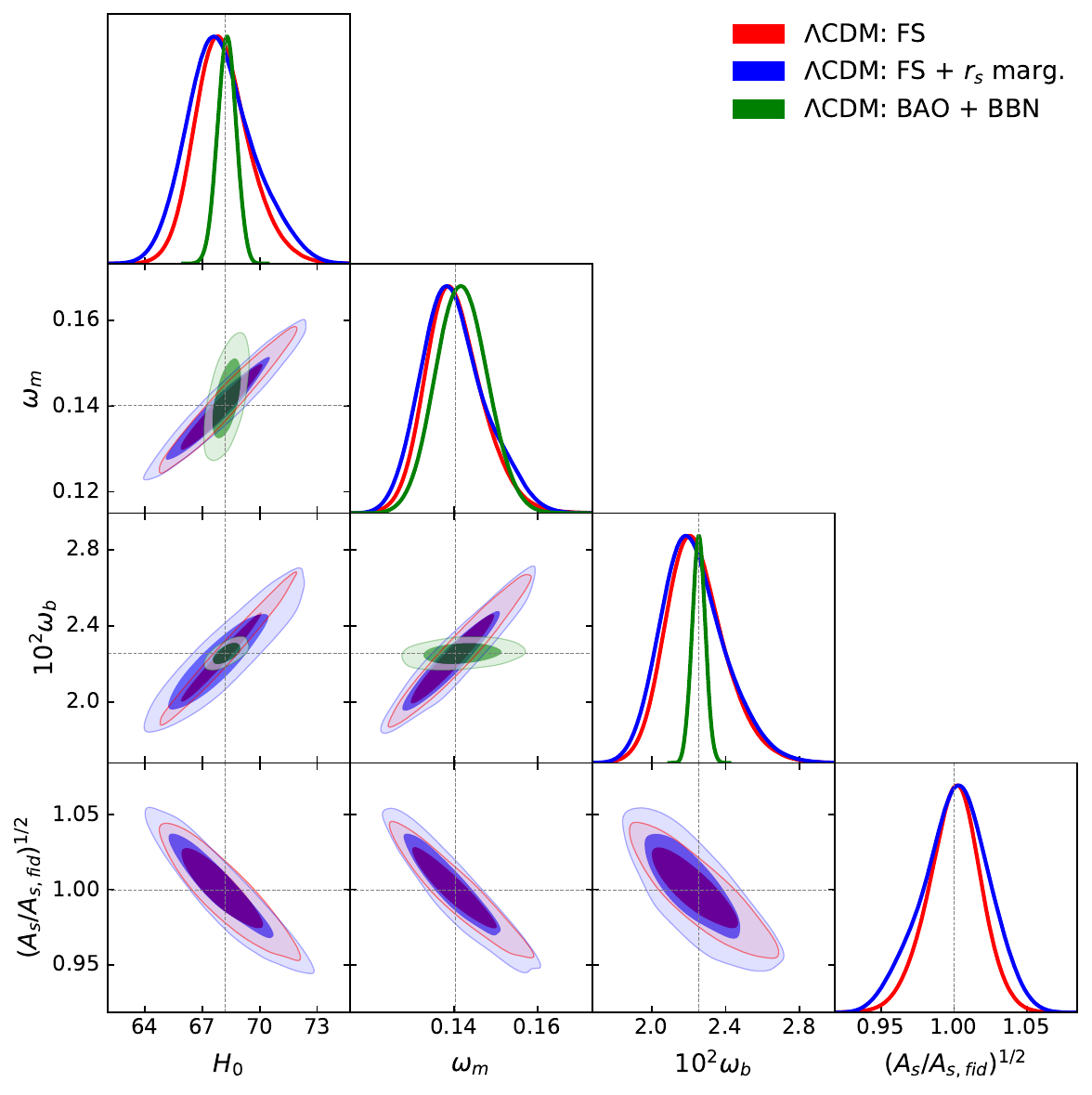}
    }
    \hfill
    \subfloat[$\Lambda$CDM: With BBN Priors]{%
      \includegraphics[width=0.48\textwidth]{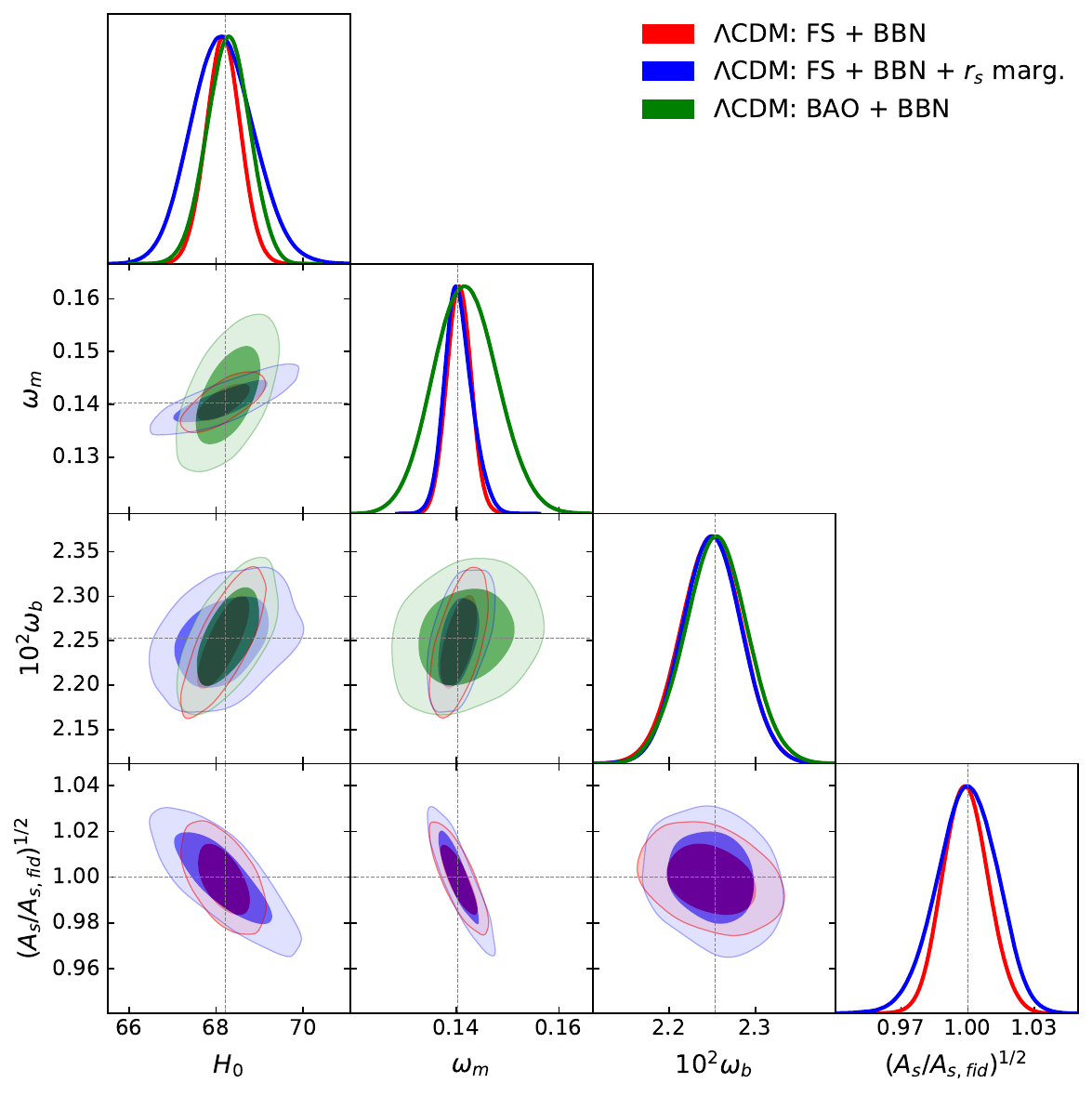}
    }
    \\
    \subfloat[\textit{Planck} + SH0ES EDE: Without BBN Priors]{%
      \includegraphics[width=0.48\textwidth]{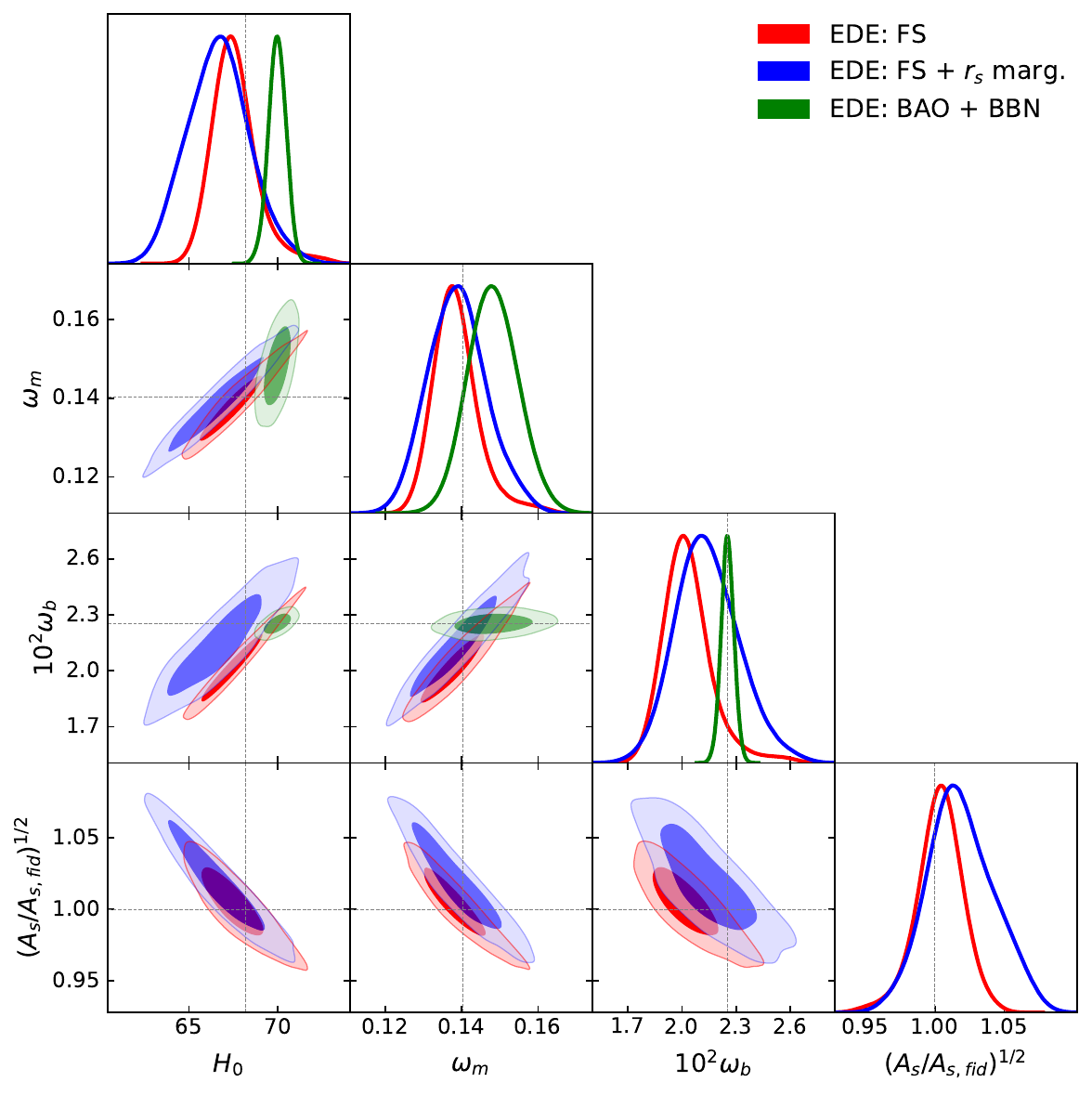}
    }
    \hfill
    \subfloat[\textit{Planck} + SH0ES EDE: With BBN Priors]{%
      \includegraphics[width=0.48\textwidth]{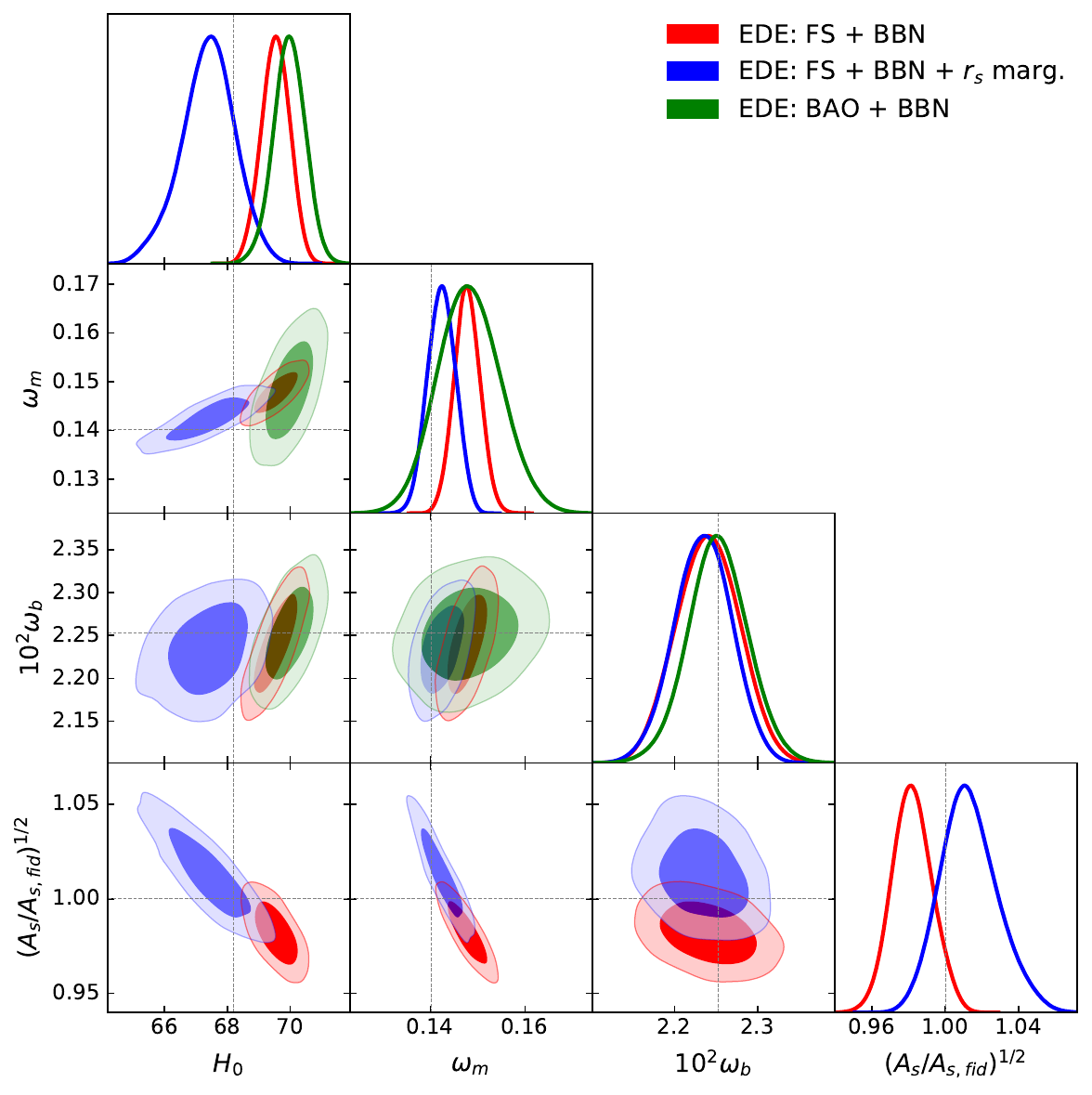}
    }
    \caption{Forecasted constraints on a Euclid-like power spectrum dataset for an underlying $\Lambda$CDM cosmology (top) or an Early dark energy cosmology (EDE; bottom), using the \textit{Planck} + SH0ES best-fit model of Ref.~\cite{Smith2020}. In both cases, data is analysed under the assumption of a $\Lambda$CDM model. Three datasets are shown: the full power spectrum likelihood (red), the full power spectrum likelihood with the addition of marginalisation over $r_s$ (blue), and the BAO-only likelihood (green). We show results both with (right panels) and without (left panels) the BBN-derived prior on the baryon density, $\omega_{\rm b} = (2.253 \pm 0.036) \times 10^{-2}$. The inclusion of baryon priors generally significantly tightens the constraints (note the changed scale of the panels between the left (w/o BBN) and right (w/ BBN) panels). The resulting $H_0$ constraints are tabulated in Tables~\ref{tab:h0-posteriors} ($\Lambda$CDM) and \ref{tab:h0-posteriors-ede} (EDE). The FS data contains information from two scales: the equality and sound horizon, which translate to different values of $H_0$ in the $\Lambda$CDM framework if the true and assumed cosmological models do not agree. In the case of EDE, including $r_s$-marginalisation shifts $H_0$ to smaller values (being equality dominated) whilst using a BAO-only likelihood shifts $H_0$ to larger values (being $r_s$-dominated); for $\Lambda$CDM we see no such shifts. When including BBN priors, the relative importance of the $r_s$ scale is amplified, shifting $H_0$ to the right for EDE, but with little change to the $r_s$-marginalised constraint. The $\approx 3\sigma$ peak shift in $H_0$ between BAO and $r_s$-marginalised FS datasets is a powerful test for internal consistency, and thus for new physics such as EDE.}
    \label{fig:oliver_plot}
\end{figure*}

Early dark energy (EDE) \citep{Karwal2016,Mortsell18,2019PhRvL.122v1301P} has recently been proposed as a potential resolution to the Hubble tension. The theory involves an additional scalar field operating in the early Universe, whose slowly rolling behavior initially mimics the phenomenology of dark energy (with equation of state $w\approx -1$), but, after the onset of field oscillations, the energy density quickly redshifts away at $z\approx \mathrm{few}\times 10^3$. By increasing the prerecombination expansion rate, this reduces the physical scale of the sound horizon, thus shifting the Hubble parameter obtained from $r_s$-dependent probes to values more compatible with direct measurements \citep[e.g.,][]{2019NatRP...2...10R}. However, to retain compatibility with the full CMB spectrum, other cosmological parameters must also shift. For example, EDE suppresses the growth of fluctuations more strongly than in $\Lambda$CDM leading to an enhanced early-integrated Sachs-Wolfe (ISW) effect in the large-scale CMB, which degrades the fit to the CMB unless the matter density $\omega_{cdm}$ is increased. Whilst such shifts are not ruled out by the CMB alone \citep{2019PhRvL.122v1301P,Hill2021}, recent papers \citep{2020PhRvD.102d3507H,Ivanov2021ede,DAmico2021} have argued that this leads to the model being disfavored when the LSS full-shape power spectrum is also included in the fit (although the statistical significance of this claim is disputed \cite{Niedermann2021b,Smith2021}\footnote{In particular, the public BOSS power spectra contained an error in the window function treatment, leading to a misnormalisation at the 10\% level \citep{deMattia19,deMattia21,beutler_mcdonald21}. As a consequence, the inferred power spectrum amplitude was about $2\sigma$ lower that that preferred from the CMB. Since there is some degeneracy between $\sigma_8$ and $f_{\rm{EDE}}$ this may also lead to a suppression in the allowed EDE fraction.}). Given the recent flurry of interest surrounding EDE, including findings that some data combinations with ACT appear to prefer this model \citep{Hill2021,Poulin2021}, it is important to consider how future LSS data can shed light on the validity of EDE solutions to the so-called $H_0$ tension.

Given that EDE is primarily active in the decade of expansion prior to recombination, it is natural to expect that its inclusion will have a different effect on the equality scale (at $z\approx 3600$), and the sound-horizon scale (at $z\approx 1090$), although the sign and magnitude of the difference is not \textit{a priori} clear. In the presence of EDE, the $H_0$ measurements obtained from the two scales (assuming $\Lambda$CDM) will not exactly agree (potentially also due to changes in other parameters required to preserve a good fit to the CMB), and hence the dataset will not be fully internally consistent. This deviation will only be present if a beyond-$\Lambda$CDM phenomenon such as EDE is active. To test this, we consider the same setup as in \S\ref{sec:test_forecast}, but instead generate the data using two different EDE cosmologies, assuming the best-fit parameters of Ref.~\cite{Smith2020} with $n=3$ (fit to \textit{Planck} and SH0ES data), and Ref.~\cite{Hill2021} (fit to ACT data), with the latter possessing a much larger EDE fraction \resub{($f_{\rm{EDE}}=0.241$ for the ACT model compared to $f_{\rm{EDE}}=0.122$ for the model fit to \textit{Planck} and SH0ES)}.\footnote{It should be noted that, while the best-fit EDE fraction in the ACT model \cite{Hill2021} is much larger, overall preference for EDE is only $\sim2\sigma$ due to large uncertainties and highly non-Gaussian posteriors. The parameters inferred from a $\Lambda$CDM fit to the same dataset also show some deviation from our fiducial model. When large scale CMB information from \textit{Planck} is added a significant preference for EDE is found but at a lower EDE fraction ($f_{\rm{EDE}}=0.113$). We select this model for illustration purposes only, showing the effect of a dramatically different cosmology on our two probes of $H_0$.} In each case, we select nuisance parameters such that the output spectrum most closely matches that of \S\ref{sec:test_forecast}. 

Data are analysed assuming the $\Lambda$CDM model, and we perform BAO analyses, FS analyses marginalised over $r_s$, and unmarginalised FS analyses. These will give $H_0$ constraints from $r_s$, $k_{\rm eq}$ and their combination, respectively. As before, we can optionally include a BBN prior on $\omega_{\rm b}$: its inclusion will strengthen both sources of $H_0$ information. Our consistency test is straightforward: does EDE induce a shift in the $H_0$ values obtained from the different standardisable rulers?

\begin{figure*}[t!]
    \subfloat[ACT EDE: Without BBN Priors]{%
      \includegraphics[width=0.48\textwidth]{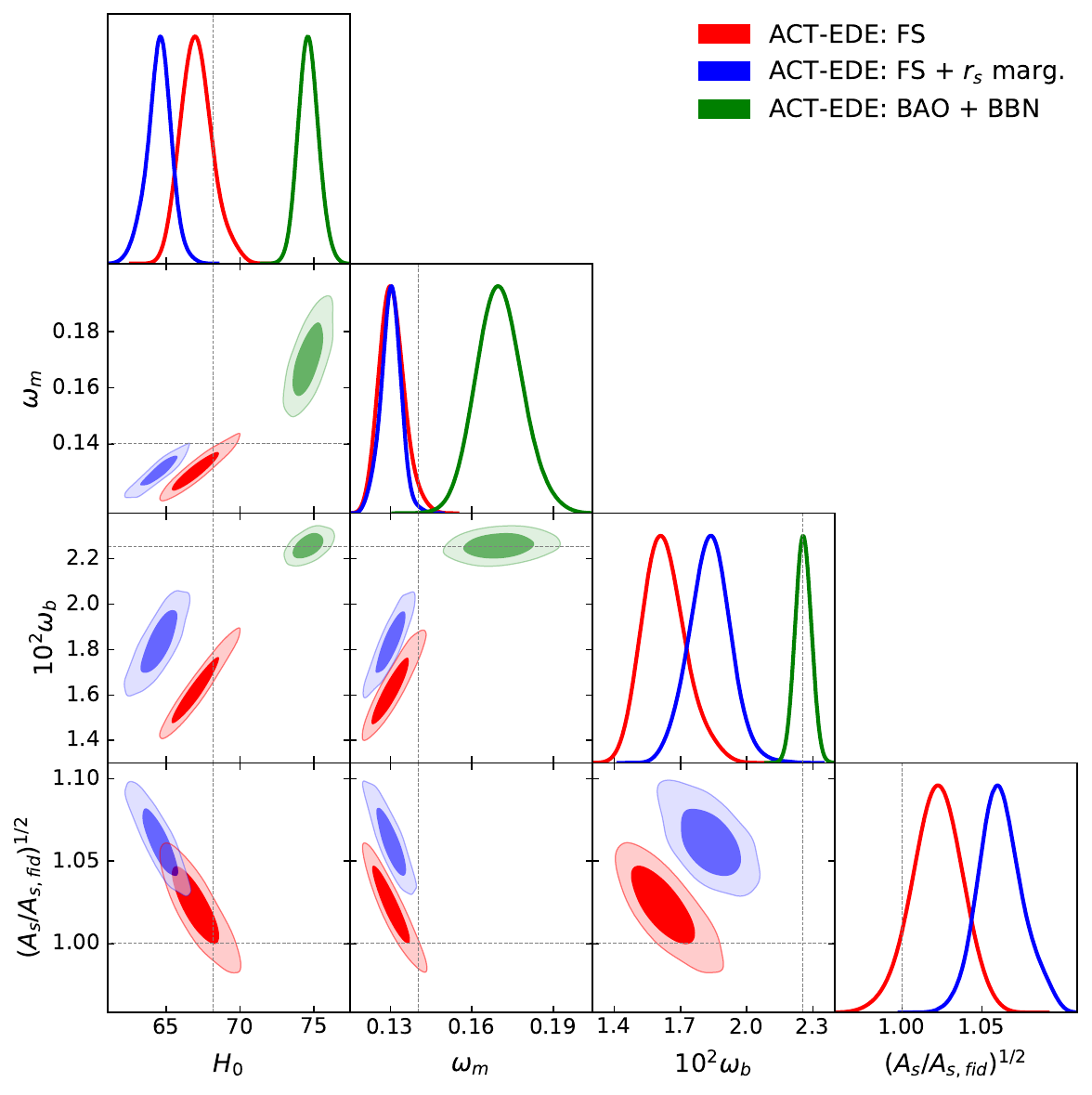}
    }
    \hfill
    \subfloat[ACT EDE: With BBN Priors]{%
      \includegraphics[width=0.48\textwidth]{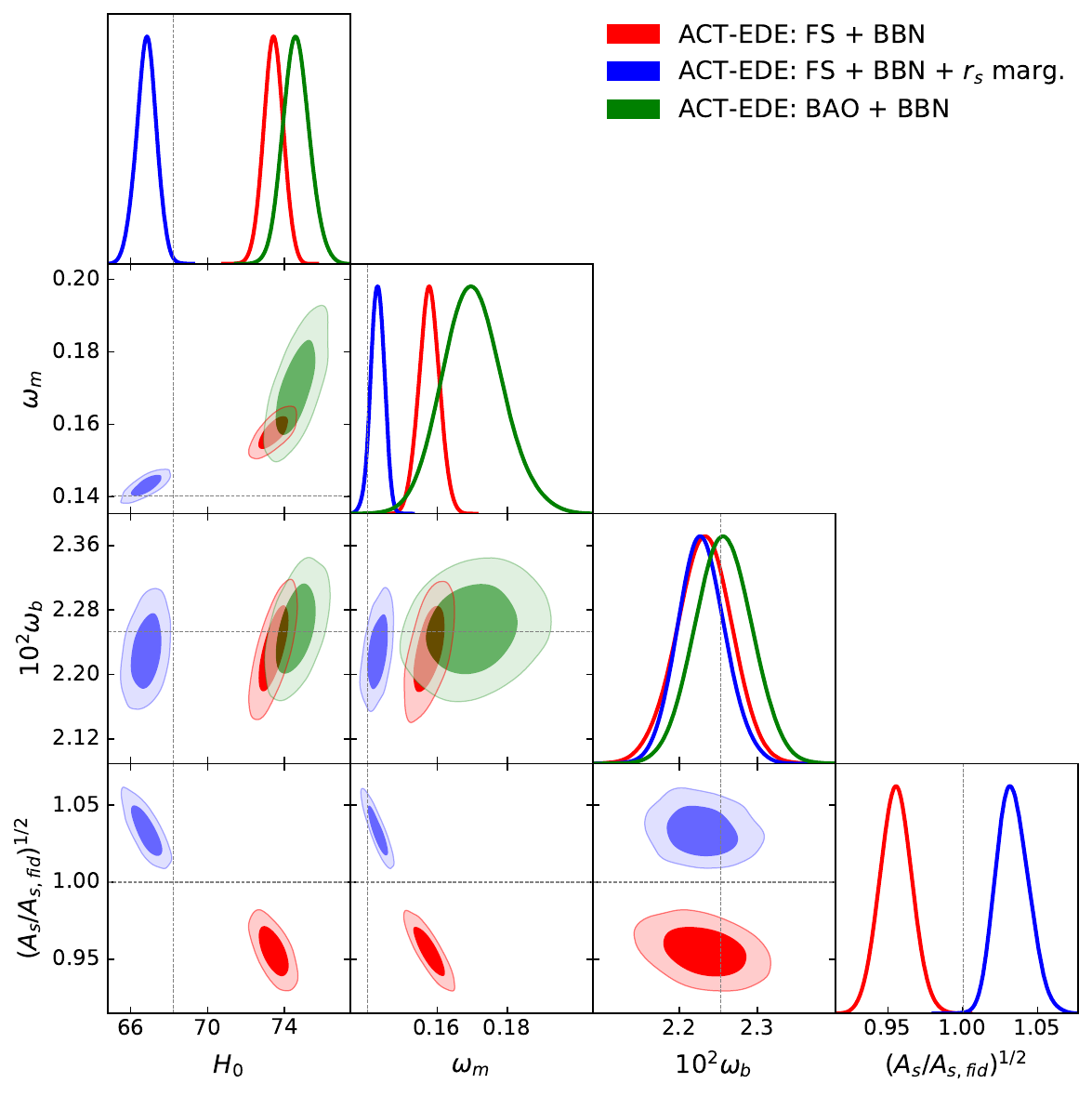}
    }
    \caption{As Fig.\,\ref{fig:oliver_plot} but assuming the best-fit EDE model from ACT data \citep{Hill2021}. In this case, the parameter shift is much more significant; we find a $9.0\sigma$ difference between $H_0$ constraints from the equality- and sound-horizon based probes.}
    \label{fig:oliver_plot-act}
\end{figure*}

In Fig.\,\ref{fig:oliver_plot} and Table\,\ref{tab:h0-posteriors-ede}, we show the parameter constraints obtained from the Euclid forecast for an EDE cosmology using the parameters of \citep{Smith2020}, analysed assuming $\Lambda$CDM. Considering first the results without BBN priors on $\omega_{\rm b}$, we observe a shift of $\Delta H_0 = 3.3\hun$ in the mean Hubble constant value between the $r_s$- and $k_{\rm eq}$-based datasets (BAO and FS + $r_s$-marginalisation respectively), which corresponds to $\sim 1.8\sigma$ with respect to the combined error bars.\footnote{This is necessarily an overestimate, since the noise in the two datasets will be correlated.} Clearly the two datasets are not fully consistent (given that the mocks do not include noise, with noise only entering into the assumed covariance matrix); this is primarily driven by the $H_0$ differences. Considering the FS (without $r_s$-marginalisation) dataset (which includes both sources of information), we find that the $H_0$ posterior lies close to that of the $k_{\rm eq}$ constraint, but is of somewhat smaller width. \resub{The differences in the $H_0$ posterior between FS and FS + $r_s$-marginalised datasets demonstrate that the BAO information still has an impact on the cosmological parameters, even when BBN priors are not imposed. This stands in contrast with the results of Ref.~\citep{Philcox2020}, which used BOSS data. We expect this to arise since the Euclid dataset has (a) larger volume, and (b) a wider redshift range. The former leads to significantly reduced errorbars compared to BOSS data, whilst the latter will constrain the matter density from shape information and coordinate distortions, allowing better internal calibration of the sound horizon.}

When a BBN prior is included in the full-shape analyses, the significance of the $H_0$ shift between $k_{\rm eq}$- and $r_s$-based measurements (i.e., BAO and FS + $r_s$-marginalisation respectively) increases to $2.6\sigma$. This is as expected: the BBN prior breaks the $H_0$-$\omega_b$ degeneracy discussed in \S\,\ref{sec:test_forecast}. In this case, the FS result (combining $k_{\rm eq}$- and $r_s$-based information) lies close to the $r_s$-only contour: this illustrates that the $H_0$ information is dominated by the sound horizon when BBN priors are used. Notably, shifts are also seen for other parameters, such as the primordial amplitude $A_s$ (which traces $\sigma_8$): this is a result of the data lacking internal consistency.

In \resub{Fig.\,\ref{fig:oliver_plot-act}} we show the analogous results using the ACT EDE model \citep{Hill2021}. In this case, the shifts are much more extreme: without a BBN prior we find $\Delta H_0 = 10.09\hun$ between the $r_s$- and $k_{\rm eq}$-based datasets, corresponding to $9.5\sigma$, and when a BBN prior is included, this becomes $\Delta H_0 = 7.8\hun$ ($9.0\sigma$). As before, the FS result (which is an average of both probes) lies close to the equality-side without BBN, but shifts to larger $H_0$ with BBN calibration. In this case, the internal tensions within the dataset are very clear; we observe shifts in a range of parameters when using the two probes, which would be clearly apparent even in the presence of observational noise. The larger shifts in this case are well understood, since the ACT model predicts an EDE fraction around twice as large as that of the \textit{Planck} + SH0ES results of \citep{Smith2020}, and consequently, a much larger change to physics at $z\sim 10^3$.

\resub{For our fiducial $\Lambda$CDM model, the comoving size of the sound horizon is $r_s=100.6 \Mpch$, while within the EDE (ACT-EDE) setup, we find $r_s=101.7\,\Mpch$ ($r_s=104.9\,\Mpch$). When performing our $r_s$-marginalised analysis (including the BBN prior on $\omega_b$) we find a best-fit sound horizon of $r_s=101.3\,\Mpch$ and $r_s=103.9\,\Mpch$ for the EDE datasets generated with the \textit{Planck}+SH0ES and ACT models respectively. For the $\Lambda$CDM model the input $r_s$ is recovered very accurately, with $r_s=100.7\,\Mpch$.\footnote{\resub{This is obtained as the product of $r_s$ computed within the best-fit $\Lambda$CDM cosmology with the corresponding $\alpha_{r_s}$ value (fitting to the simulated data while allowing for sound horizon rescaling and including the usual prior on $\omega_b$).}} The equality scale within the $\Lambda$CDM model, computed as the comoving size of the horizon at matter-radiation equality, is $k_{\rm{eq}}=0.015 \hMpc$. In the context of an EDE cosmology, early dark energy makes an additional non-negligible contribution to the background evolution of the universe around matter-radiation equality. Consequently, mode growth pre- and post-equality are modified such that we do not necessarily expect the turnover scale of the matter power spectrum to correspond to the size of the horizon at matter-radiation equality. Computing $k_{\rm{eq}}$ na\"ively for the \textit{Planck}+SH0ES and ACT models we find $k_{\rm{eq}}=0.011\,\hMpc$ and $k_{\rm{eq}}=0.009\,\hMpc$ respectively. When considering the best-fit $\Lambda$CDM models from our marginalised analysis again we find $k_{\rm{eq}}=0.015\,\hMpc$ and $k_{\rm{eq}}=0.015\, \hMpc$, which is very similar to our fiducial $\Lambda$CDM model (for which the best-fit cosmology also yields $k_{\rm{eq}}=0.015\,\hMpc$).}

The conclusion of this exercise is clear: modifications to early Universe physics affect the two available standard rulers, the sound horizon scales and the equality scale, differently and (potentially in conjunction with changes in the other relevant parameters) can give a measurable shift in the $H_0$ parameter inferred from each within \lcdm. For the best-fit model of \citep{Smith2020}, the shifts are modest, but marginally detectable, whilst for that of \citep{Hill2021}, they appear at high significance. 
Furthermore, whilst the above exercise has been performed in the context of EDE, this consistency test is much more general: although the signal-to-noise obtained will be survey and model dependent, any model of new physics that modifies the sound horizon and equality scale differently could, in principle, be detected by comparing the $k_{\rm eq}$- and $r_s$-derived constraints.


\section{Application to BOSS}\label{sec: boss}
Previous attempts to compute $H_0$ from the equality scale \citep{2020arXiv200704007B,Philcox2020} reduced dependence on $r_s$ by removing the commonly applied BBN prior, and thus the BAO calibration. In the above, we have demonstrated that the addition of BBN information on $\omega_{\rm b}$ can sharpen the $k_{\rm eq}$ standard ruler, if it is additionally combined with $r_s$-marginalisation techniques to remove any BAO- or sound-horizon-derived information. This motivates the question: can such an approach be used to strengthen the $r_s$-independent $H_0$ constraint from current data?

To answer this, we repeat the analysis of Ref.~\citep{Philcox2020}, utilising the former work's public likelihoods.\footnote{Available at \href{https://github.com/oliverphilcox/montepython_equality}{GitHub.com/oliverphilcox/montepython\_equality}.} These are similar to those of the above Euclid analysis, but additionally include the effects of the BOSS survey geometry, \resub{fix $k_{\rm min}=0.01\hMpc$ and $k_{\rm max}=0.25\hMpc$, using two redshift slices at $z = 0.38$ and $0.61$.}\footnote{Note that, as previously mentioned, the public BOSS power spectra contained an error in the window function treatment, leading to a misnormalisation at the $10\%$ level \citep{deMattia19,deMattia21,beutler_mcdonald21}. Whilst this effect is important for inferences involving $\sigma_8$, it is not expected to affect the $H_0$ constraints.} Unlike the former work, we remove the Gaussian priors on $A_s$, $\omega_{\rm b}$ and $\Omega_m$ (replacing them by broad flat priors); furthermore, we modify the likelihoods to allow for $r_s$-marginalisation, using the approach of \S\ref{sec: method}. In this form, our likelihoods can assess the information content of the BOSS power spectrum alone, and can be optionally combined with $\omega_{\rm b}$ priors from BBN \citep[e.g.,][]{2019JCAP...10..029S} and $\Omega_m$ priors from Pantheon supernovae \citep{2018ApJ...859..101S}. \resub{For comparison, we consider the information present solely in the sound-horizon feature, by performing a BAO-only analysis using a Markov chain analysis of the Alcock-Paczynski parameters measured in \citep{2020JCAP...05..032P}. This is similar to the BAO-only analysis discussed in \ref{subsec: validation}, but includes the effects of BAO reconstruction, which enhances the information content.}

\begin{figure}[t!]
    \centering
    \includegraphics[width=0.5\textwidth]{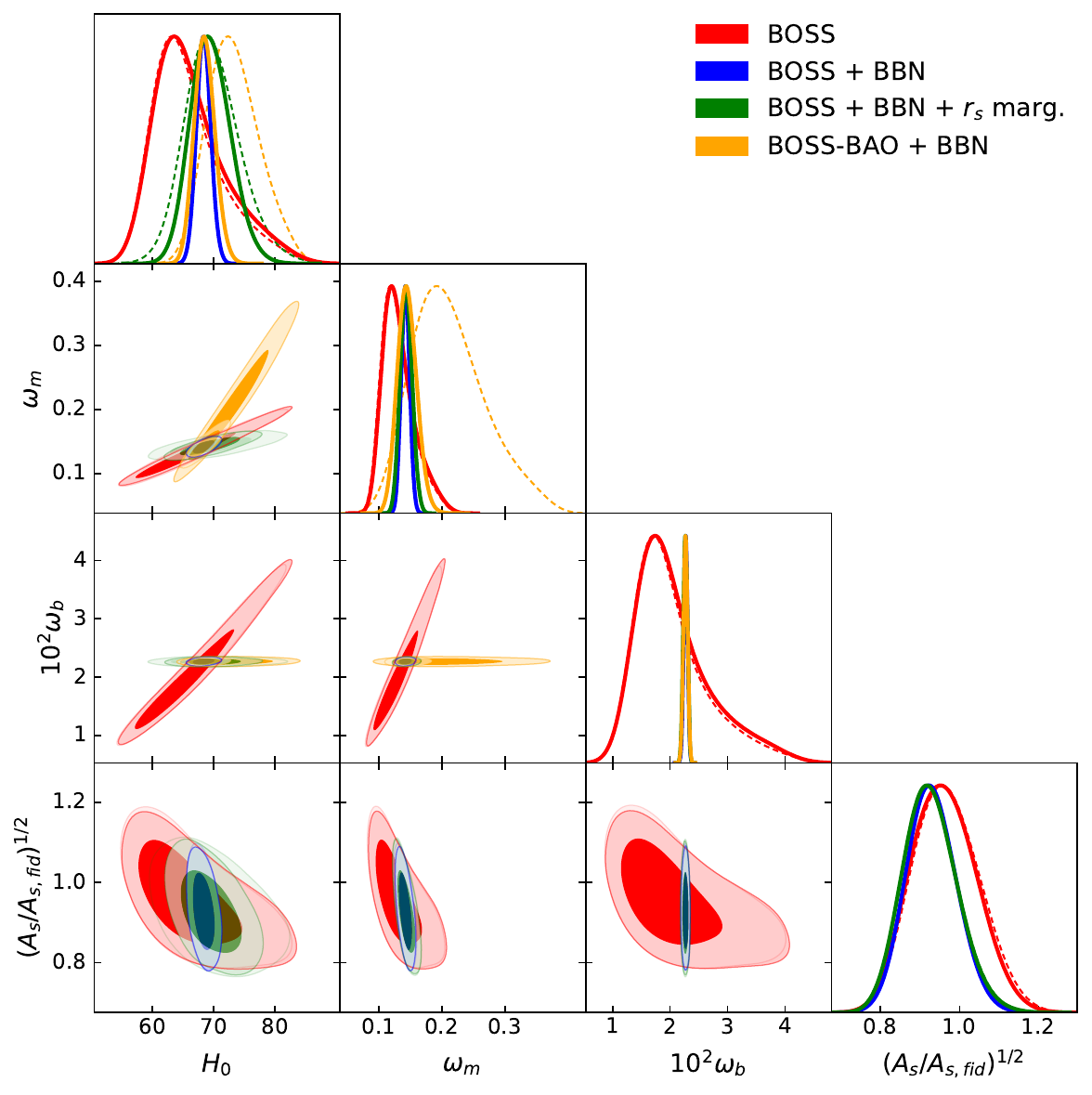}
    \caption{Cosmological constraints obtained from the BOSS DR12 data-set, in combination with BBN priors on $\omega_{\rm b}$ and Pantheon priors on $\Omega_m$. The BOSS-only constraints (red) are similar to those of \citep{Philcox2020}, while those including BBN (blue) are similar to those of \citep{Ivanov2017}. \resub{We additionally show BAO-only results (yellow), obtained following \citep{2020JCAP...05..032P}.} The addition of $r_s$-marginalisation (green, which is the principal new feature of this work) allows BBN priors to be included in the analysis without incurring dependence on $r_s$; this shrinks the parameter contours considerably compared to the BOSS-only results. The primary contours include Pantheon priors on $\Omega_m$: in dotted lines and faint contours, we show the results without this prior. Notably, the prior has little effect, except for slightly shrinking the $H_0$ and $\Omega_m$ contours after $r_s$-marginalisation, \resub{and significantly tightening the BAO-only results}. $1\sigma$ $H_0$ constraints for this sample are given in Table\,\ref{tab:h0-boss}.}
    \label{fig:boss-results}
\end{figure}

The results of this analysis are shown in Fig.\,\ref{fig:boss-results}, with $H_0$ posteriors given in Table\,\ref{tab:h0-boss}. For the BOSS + Pantheon analyses, our results may be compared to \citep{Philcox2020}, which found $65.1^{+3.0}_{-5.4}\hun$; here we find $66.1^{+3.8}_{-7.0}\hun$, with the slight broadening linked to the removal of broad $\omega_{\rm b}$ and $A_s$ priors. For the BOSS + BBN analysis, which does depend on the sound horizon, the analogue is \citep{Ivanov2017}; this analysis obtained $67.9\pm1.1\hun$, similar to the results found herein. \resub{Finally, the BOSS BAO-only analysis may be compared to Fig.\,2 of \citep{2019JCAP...10..029S}.} Of greatest interest are the constraints including both BBN and $r_s$-marginalisation, i.e. those utilising the new methods developed in this work. As shown in the figure, the constraints including both $r_s$-marginalisation and priors on $\omega_b$ are narrower than those obtained from the BOSS data alone, and the distribution is significantly closer to Gaussian. In both cases, we exclude sound-horizon information: first, explicitly, and second, by removal of the BBN prior. The addition of marginalisation and BBN information shrinks the $H_0$ posterior by $\approx 0.8\hun$ ($H_0 = 69.8^{+3.9}_{-4.9}\hun$) relative to the BOSS-only information ($H_0 = 65.6^{+3.6}_{-6.7}\hun$), or $\approx 2.2\hun$ if Pantheon priors on $\Omega_m$ are included ($H_0 = 69.5^{+3.0}_{-3.5}\hun$ for `FS+BBN+$r_s$ marg.+Pantheon' compared to $H_0 = 66.1^{+3.8}_{-7.0}\hun$ `FS+Pantheon').
\resub{Without Pantheon priors, the marginalised full shape analysis is comparable to the BAO-only analysis. However, inclusion of prior information on $\Omega_m$ from Pantheon leads to improved calibration of the sound-horizon feature, giving a tight constraint with $\sigma_{H_0}\approx 1.7\hun$ from BAO alone, only somewhat wider than the FS + BBN results.}

Additionally, we find a rise in the central $H_0$ value by $\sim 3\hun$ and $\sim 4\hun$ when BBN information is added in the analysis with and without priors on $\Omega_m$ respectively. However, this is not unexpected given that new BBN data is added and the error bars change significantly, and so (as also suggested by a simple estimate\footnote{Following the prescription from Ref.\,\cite{Gratton2020} we estimate the significance of these shifts at $\sim$\,$1.6\sigma$ and $\sim$\,$0.8\sigma$ respectively. For this estimate we have taken half the posterior width as our estimate of the $1\sigma$ constraint on $H_0$.}) we do not affix any significance to this change; \resub{furthermore it is consistent with the BAO-only dataset}. Notably, the increase in $H_0$ precision from adding BBN information (and marginalising over $r_s$) is less than would be expected from rescaling the Euclid results of \S\ref{sec:test_forecast} to the BOSS effective volume. This can be understood by noting that the BBN prior on $\omega_{\rm b}$ has the same fractional width in both scenarios. Given the smaller survey volume of BOSS, the BAO information is expected to have a greater impact, thus the equality constraints are fractionally weaker.

In the previous sections, we have demonstrated that constraints obtained using $r_s$-marginalisation are sound-horizon independent for a Euclid-like survey, even when a BBN prior is applied. For BOSS, the same conclusion should naturally hold, since its volume (and thus the precision with which the equality and sound-horizon features can be measured) is much smaller. Our constraints, therefore, represent the tightest sound-horizon-independent bounds on $H_0$ from current galaxy surveys; including both BBN and Pantheon data, this has the value $H_0 = 69.5^{+3.0}_{-3.5}\hun$, \resub{fully consistent with the BAO-only constraints of $68.5^{+1.6}_{-1.8}\hun$ (with the $r_s$ prior set by BBN and Pantheon).} Whilst this result is not yet able to shed light on the so-called Hubble tension, it is a significant sharpening of previous constraints, and, as we have shown, such constraints are expected to tighten significantly with future data.


\begin{table}[]
    \caption{$H_0$ constraints from the BOSS DR12 full-shape (FS) data-set, optionally combined with priors on $\Omega_m$ (from Pantheon supernovae) and $\omega_{\rm b}$ (from BBN). The respective contours are shown in Fig.\,\ref{fig:boss-results}.\label{tab:h0-boss}}
    \centering
    \begin{tabular}{l|c|c}
    \hline\hline
     & BOSS &BOSS + Pantheon \\\hline
     FS & $65.6^{+3.6}_{-6.7}$ & $66.1^{+3.8}_{-7.0}$\\
     FS + BBN & $68.3\pm 1.2$ & $68.3\pm 1.2$\\
     FS + BBN  + $r_s$ marg. & $69.8^{+3.9}_{-4.9}$ & $69.5^{+3.0}_{-3.5}$\\
     \resub{BAO + BBN} & \resub{$73.2^{+3.7}_{-4.8}$} & \resub{$68.5^{+1.6}_{-1.8}$}\\\hline\hline
    \end{tabular}
\end{table}


\section{Conclusion}\label{sec: conclusion}

Obtaining accurate measurements of the Hubble parameter is a key goal for current and future galaxy surveys. In this work, we have demonstrated that a full shape (FS) analysis of forthcoming Euclid spectroscopic data has access to two powerful and physically independent probes of $H_0$. When the analysis is performed with priors on the baryon density from BBN, the physical size of the sound horizon, $r_s$, is calibrated and this information dominates the $H_0$ constraint. Without prior information on $\omega_b$ the $H_0$ constraints are instead dominated by the matter-radiation equality scale, $k_{\rm eq}$, which is sensitive to higher-redshift physics. Here we have shown that, using a heuristic procedure to marginalise over the size of the sound horizon, we are able to avoid $r_s$-based information entering our analysis even when including a prior on $\omega_{\rm b}$. This procedure leads to equality-based constraints on $H_0$ that are competitive with those derived from $r_s$-based probes. Testing our method with an MCMC analysis of a mock Euclid likelihood, we forecast $\sigma_{H_0}=0.72\hun$ with this new method compared to $\sigma_{H_0}=0.49\hun$ from BAO information and $\sigma_{H_0} = 1.4\hun$ for a full shape analysis without any external baryon information (in the manner suggested in previous works).

As discussed for example in Ref.~\cite{Ross2012}, systematic uncertainties on large scales arising for example from a spatially varying targeting efficiency on the sky (caused by various observational challenges such as atmospheric transparency, sky brightness or foreground contaminants) remain a major challenge for LSS surveys. It has been shown that while BAO observations are relatively insensitive to these systematics the broadband power spectrum can be significantly affected by spurious residuals, especially on large scales  \cite{Ross2012}. To perform a measurement as proposed in this work with future surveys it will be essential to demonstrate carefully that all relevant observational systematic effects are under control. Various mitigation strategies have been proposed for this purpose (see e.g. Ref.~\cite{Weaverdyck2021}). We defer a detailed investigation of systematic challenges to our method (and their mitigation) to future work.

In \S\ref{sec:test_forecast} we have demonstrated that, when analyzing a mock dataset generated with a \lcdm cosmology, different analysis methods (FS, FS+BBN, FS+BBN+$r_s$ marg. and BAO+BBN) yield consistent constraints on $H_0$, as expected. When the data is instead generated from an EDE model, but analysed assuming \lcdm, we find measurable tensions between the $H_0$ constraints inferred from FS+BBN+$r_s$ marg. and BAO+BBN analyses (see \S\ref{sec: new-models}). This occurs since a period of early dark energy domination affects the sound horizon and equality scales differently.

For the \textit{Planck}+SH0ES best-fit model of \citep{Smith2020}, the shift in $H_0$ falls just short of being detected at $3\sigma$; however, if we instead use that of the latest ACT analyses \citep{Hill2021}, the shift increases to $\approx 9\sigma$, due to the much larger EDE fraction. The results found herein demonstrate that our method can be used as a powerful null test for models of post-\lcdm physics. Since many approaches claiming to resolve the so-called Hubble tension modify the physical size of the sound horizon at the end to the baryon drag epoch, it is generically expected that such models would affect the sound horizon and the equality scale differently, thus leading to an inconsistency of the two $H_0$ measurements.

As mentioned previously, a multitude of different models have been proposed to resolve the so-called $H_0$ tension. Whilst the most stringent constraints on each of these models are certainly derived from a dedicated analysis, the effort necessary to probe (and wherever possible rule out) a large number of models as well as mechanisms that have not yet been proposed, makes it useful to generically test all $r_s$-modifying models. With future surveys this approach will allow one to gauge how promising such modifications are as resolutions to the Hubble tension. A fully parametric model to fit the full shape power spectrum (as in Ref.~\cite{Brieden2021}) could provide a similar generic probe. However, there are in principle many ways to devise such a parametrisation, and hence we here chose a physically motivated approach that explicitly aims to decouple the two relevant sources of information contained within the power spectrum. It will be exciting to see the results of such studies applied to future data.

\begin{acknowledgments}
\footnotesize
We thank Graeme Addison, Eric Baxter, Florian Beutler, Simone Ferraro, Hector Gil-Marin, Colin Hill, Misha Ivanov, Adam Riess, Tristan Smith and Zvonimir Vlah for useful discussion and comments. \resub{We are additionally grateful to the anonymous referee for insightful comments.} GSF acknowledges support through the Isaac Newton Studentship and the Helen Stone Scholarship at the University of Cambridge. OHEP acknowledges funding from the WFIRST program through NNG26PJ30C and NNN12AA01C and thanks the Simons Foundation for additional support. 

The authors are pleased to acknowledge that the work reported in this paper was substantially performed using the Helios cluster at the Institute for Advanced Study. Additional computations were performed using the Princeton Research Computing resources at Princeton University which is a consortium of groups led by the Princeton Institute for Computational Science and Engineering (PICSciE) and the Office of Information Technology's Research Computing Division.

\end{acknowledgments}
\onecolumngrid

\appendix
\section{Broadband $r_s$ suppression}\label{app:rs_braodband_suppression}

\begin{figure}[t!]
    \centering
    \includegraphics[width=0.5\textwidth]{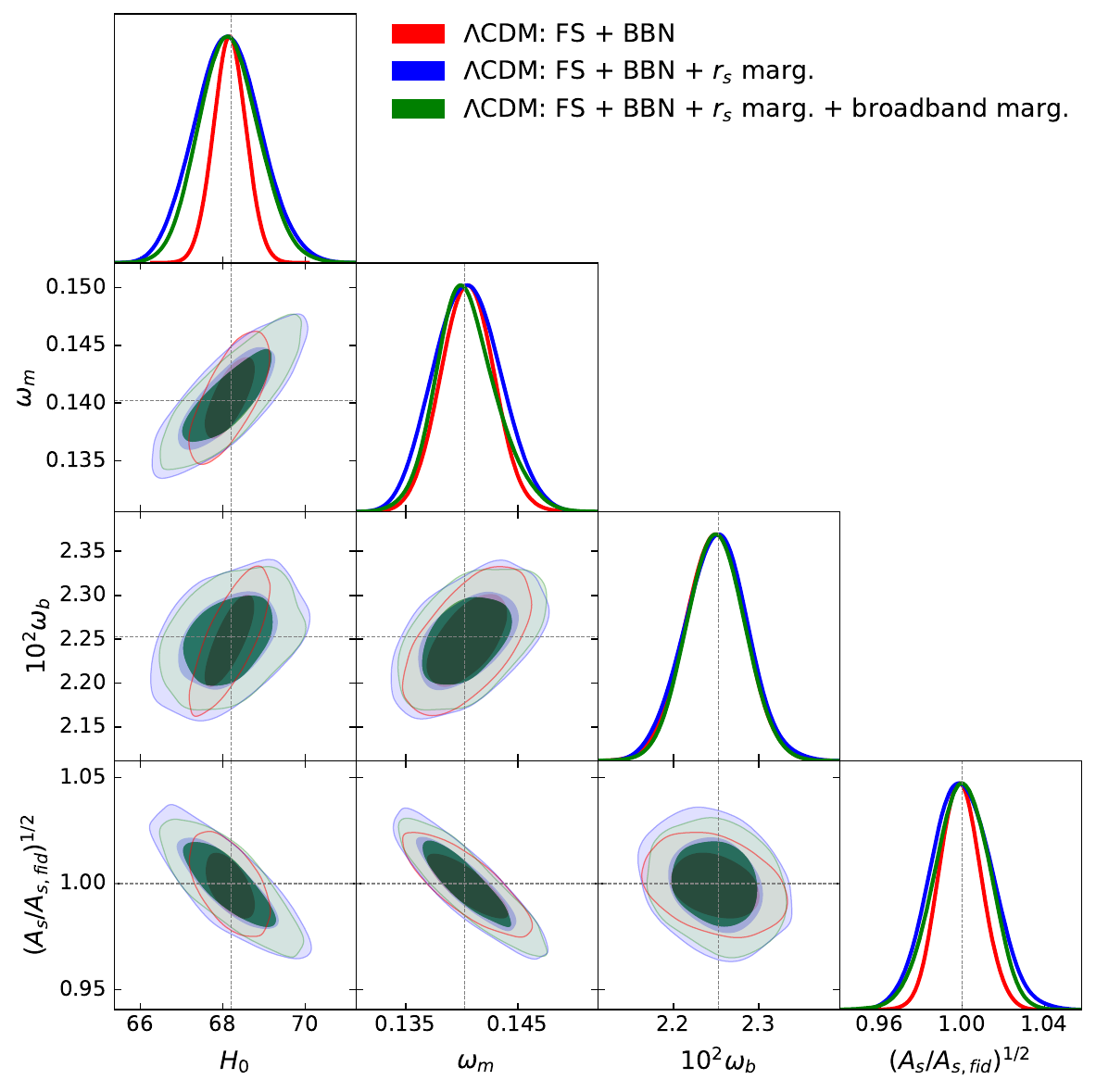}
    \caption{$\Lambda$CDM parameter constraints including marginalisation over both the sound horizon, $r_s$, and the baryon-induced broadband suppression. This is implemented using the marginalisation matrix described in Appendix \ref{app:rs_braodband_suppression}. Notably, the additional marginalisation over the baryon-induced broadband suppression 
    does not appreciably affect the constraints, indicating that the simple $r_s$-marginalised likelihood we used is already free of sound horizon information. The $H_0$ constraints are $H_0  = 68.17\pm 0.40	$, $68.15\pm 0.72$ and $68.12\pm 0.79$ respectively (in $\hun$ units).} 
    \label{fig:broadband-marg}
\end{figure}

As mentioned in \S\ref{sec: method}, the BAO feature is not the only source of information about $r_s$ (and thus $H_0$). Additionally, the sound horizon at photon-baryon decoupling also affects the scale on which baryonic effects suppress structure formation. This effect is typically relevant on scales similar to the BAO scale. Here we approximately marginalise over a shift in the baryon suppression scale by modifying the data covariance as described in \S\ref{sec: method}. We conservatively assume that any rescaling of the baryon suppression scale is uncorrelated with other variations in the nonlinear power spectrum.

The results of analysing the mock Euclid data using the modified covariance matrix are shown in Fig.\,\ref{fig:broadband-marg}. We observe almost no difference between constraints derived with and without the broadband $r_s$-marginalisation. In \S\ref{sec:test_forecast}, we demonstrated that our $H_0$ constraints were independent of the sound horizon scale, even when neglecting the impact of $r_s$ on the broadband power spectrum; in concert with the above, this indicates that the effect of broadband $r_s$-marginalisation for our Euclid analysis is negligible. We also verified this using a Fisher forecast run with an analytical Eisenstein-Hu \cite{1998ApJ...496..605E} transfer function in which we explicitly include marginalisation over the parameter $\beta_{r_s}$ which rescales the suppression scale as introduced in \S\ref{sec: method}. We find no appreciable impact on cosmological parameter constraints. This analysis also indicates that changes in the baryon suppression scale are strongly degenerate with various bias and counterterm parameters. Both tests are sensitive only to the linear-order changes in the power spectrum induced by baryon suppression. However, we find that the higher-order derivatives of the power spectrum with respect to $\beta_{r_s}$ are negligible compared to the first, and hence we do not believe higher-order effects to be important.

\section{Constraints on $H_0$ from power spectra without the BAO feature}\label{app:no_wiggle}
\begin{figure}[!t]
    \centering
    \includegraphics[width=0.7\textwidth]{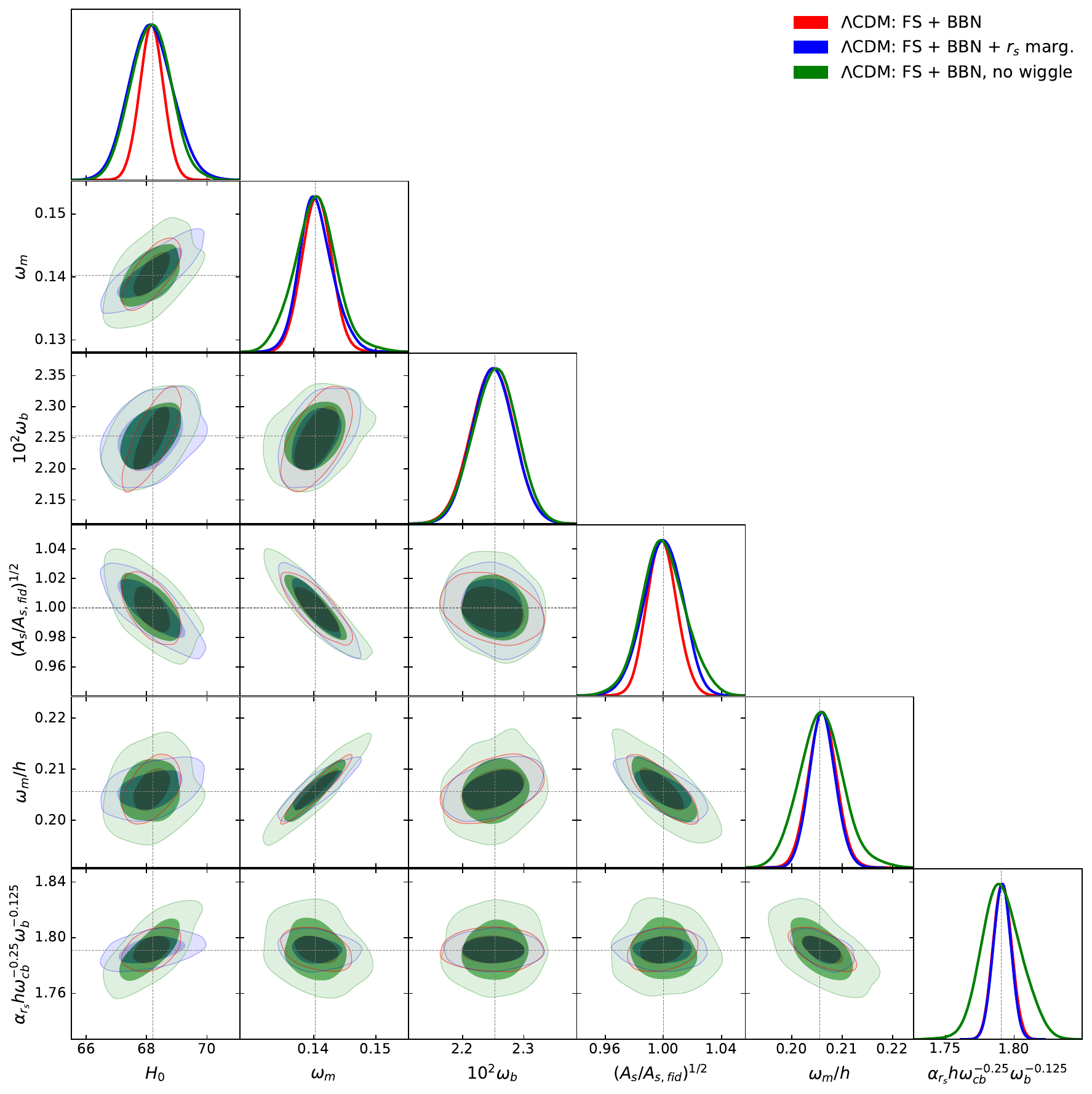}
    \caption{Comparison of the cosmological parameter constraints from a Euclid $\Lambda$CDM power spectrum forecast to those obtained from a `no-wiggle' power spectrum (green) (including a BBN derived prior on $\omega_b$ in all cases). The latter case contains no oscillations, and thus no information about the sound horizon, except that in the power spectrum broadband (shown to be insignificant in Appendix \ref{app:rs_braodband_suppression}). The $H_0$ constraints from the `no-wiggle' model are similar to those from the $r_s$-marginalised analysis, but significantly weaker than the unmarginalised constraints, showing that even without any information from the BAO feature (excluding not only the spacing of peaks but also their amplitudes) meaningful constraints on $H_0$ can be extracted from the broadband power spectrum alone. The $H_0$ constraints are $H_0  = 68.17\pm 0.40$, $68.15\pm 0.72$ and $68.15\pm 0.66$ (in $\hun$ units), for the `FS+BBN', `FS+BBN+$r_s$ marg.' and `FS+BBN (no wiggle)' analyses respectively. (Note: The constraint on $h \omega_{cb}^{-0.25}\omega_{\rm b}^{-0.125}$ from the `no-wiggle' dataset is not indicative of residual sound horizon information but rather, since no rescaling is allowed, a consequence of broadband derived constraints on $h$, $\omega_{cb}$, and $\omega_{\rm b}$.)}
    \label{fig:lcdm_wiggle_plot}
\end{figure}

In Fig.\,\ref{fig:lcdm_wiggle_plot} we show a comparison of parameter constraints from our FS forecasts to those obtained from a `no-wiggle' power spectrum (including a BBN derived prior on $\omega_b$ in all cases). Such a `no-wiggle' dataset is generated by setting the wiggly part of the power spectrum, $P_w$ (obtained as described in \S\ref{sec: method}), to zero within \texttt{CLASS-PT}. We observe that while the $H_0$ constraint from the `no-wiggle' analysis is significantly wider than the one from the `FS+BBN' analysis, it matches that of the `FS+BBN+$r_s$ marg.' analysis well ($H_0=68.15\pm 0.66 \hun$ compared to $H_0=68.15\pm 0.72\hun$ respectively). This shows that meaningful constraints on $H_0$ can be derived from the broadband alone. However, constraints on a variety of other parameters, including for example $\omega_{m}$, differ between the `no-wiggle' and the $r_s$-marginalised analyses. This indicates that, as discussed in the body of this work, there is residual, non-$r_s$-related, information contained in the `wiggly' part of the power spectrum even when rescaled, such as the information on the baryon to dark matter density ratio encoded in the BAO peak heights. The good agreement between $H_0$ constraints is likely explained by a tradeoff between this additional information and the additional uncertainty in the broadband shape introduced by adding and shifting the `wiggly' component.

\section{Analysis with varying $k_{\rm{max}}$ and redshift selections}\label{app:kmax_z_vary}

\resub{Here we consider different subsets of our data to explore the origin of the information content in our analysis. This is performed using Fisher forecasts for a Euclid-like survey, which were shown to be in excellent agreement with the MCMC analysis in \S\,\ref{subsec: rs-tests}.} 

\begin{figure}
    \centering
    \includegraphics[width=0.7\textwidth]{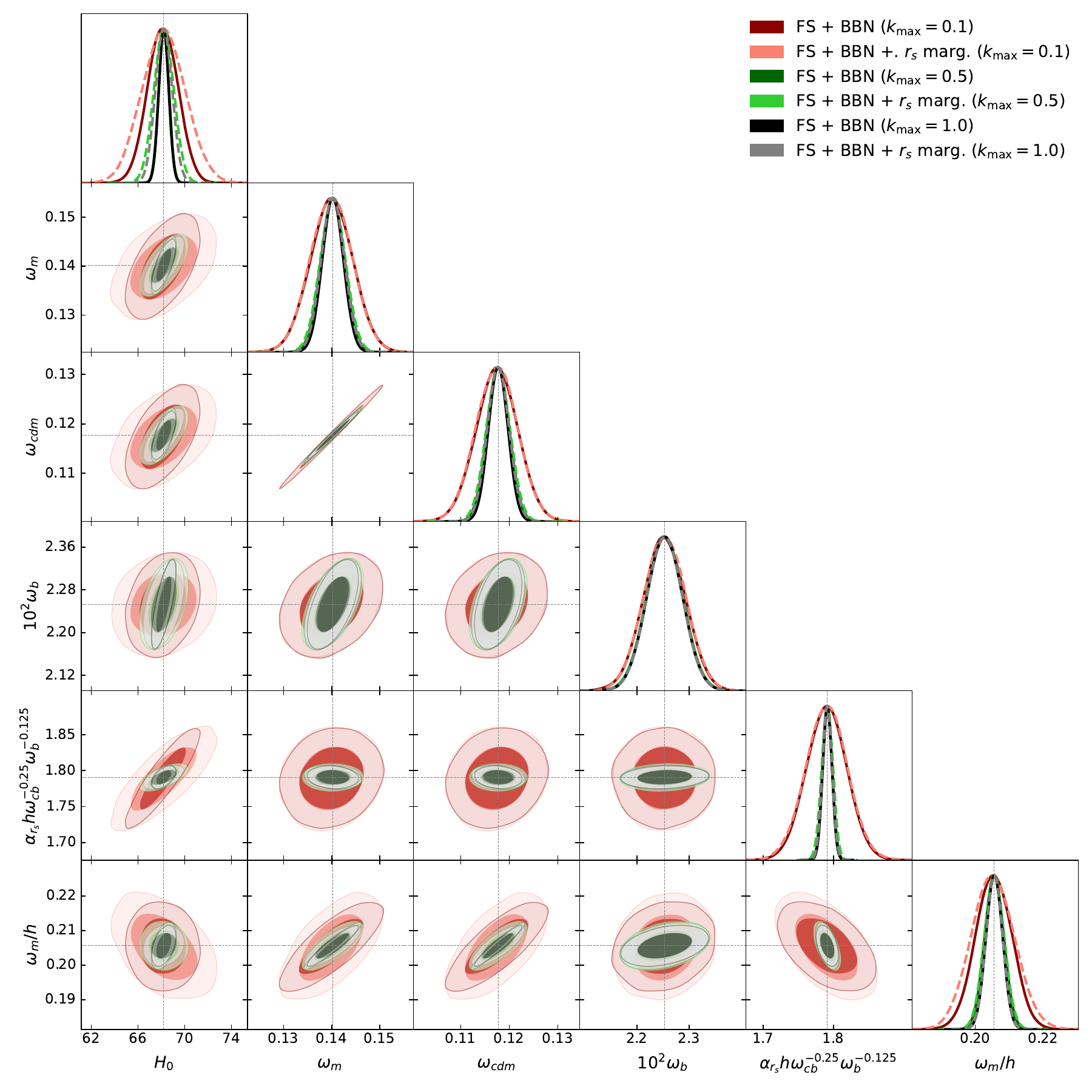}
    \caption{\resub{Fisher forecast of $\Lambda$CDM parameter constraints for a a Euclid-like survey for different choices of $k_{\rm{max}}$. The marginalised constraints on $H_0$ are given in Table~\ref{tab:kmax_z_forecast}. Marginalisation over the sound horizon significantly degrades constraints when only information below $k=0.1\hMpc$ is used, primary due to a lack of $\omega_m$ information from the power spectrum at higher $k$. We further note that, even though only minimal $r_s$-information is present below $k=0.1\hMpc$, one still expects $r_s$-marginalisation to degrade $H_0$ constraints. This will occur since the the first BAO peak partially overlaps with the turnover scale of the power spectrum, implying that $r_s$-marginalisation will decrease the precision with which $k_{\rm{eq}}$ can be determined. The additional information content on scales $k\gtrsim0.5\hMpc$ is small due to large theoretical errors included in the covariance.}}
    \label{fig:kmax_forecast_plot}
\end{figure}

\resub{First, we consider a range of different scale cuts, setting the maximum wavenumber in the analysis to $k_{\rm{max}}=0.1$, $0.5$ or $1.0 \hMpc$ (the latter being the default value in the above analysis). As mentioned in \S\ref{subsec: pk-model}, our covariance matrix accounts for theoretical errors due to higher order loop corrections (see \S\,\ref{sec:test_forecast} and Ref.~\cite{Chudaykin2019}). This leads to the errors beyond $k\sim 0.3\hMpc$ increasing dramatically and the data becoming relatively uninformative on small scales even for the lowest redshift bins where shot noise is subdominant until $k\approx 0.5\hMpc$ (see Table~\ref{tab:nuisance_params}). We thus expect only minor differences between $k_{\rm{max}}=0.5\hMpc$ and $k_{\rm{max}}=1.0\hMpc$, matching the Fisher forecast results of Fig.\,\ref{fig:kmax_forecast_plot}. The figure also shows that, when $r_s$-marginalisation is included, constraints are degraded for all three choices of $k_{\rm{max}}$, by a factor of $\sim$1.5, $\sim$2.0 and $\sim$1.7 for $k_{\rm{max}}=0.1$, $0.5$, and $1.0\hMpc$ respectively (cf.\,Table~\ref{tab:kmax_z_forecast}). Whilst there is little sound horizon information contained below $k=0.1\hMpc$, allowing the BAO scale to shift freely also degrades $k_{\rm{eq}}$-derived constraints on $H_0$. This occurs because the first BAO peak partially overlaps with the turnover scale of the power spectrum and $r_s$-marginalisation therefore decreases the precision with which $k_{\rm{eq}}$ can be determined from the power spectrum.}

\resub{Our $r_s$-marginalised constraints are improved by a factor of $\sim$2.2 when increasing our $k_{\rm{max}}$ from $0.1$ to $0.5\hMpc$; the unmarginalised constraints are improved by a factor of $\sim$3. While this is in contrast to our previous work \cite{Philcox2020} where we saw little improvement in the equality derived constraints when increasing $k_{\rm{max}}$ beyond $0.1\hMpc$, it is expected since with our new method we are able to use some of the information contained in the BAO feature as discussed previously while explicitly removing the information on the sound horizon scale. One also expects the more dramatic improvement in the unmarginalised constraints since essentially all the BAO information falls within this range.}

\begin{figure}
    \centering
    \includegraphics[width=0.7\textwidth]{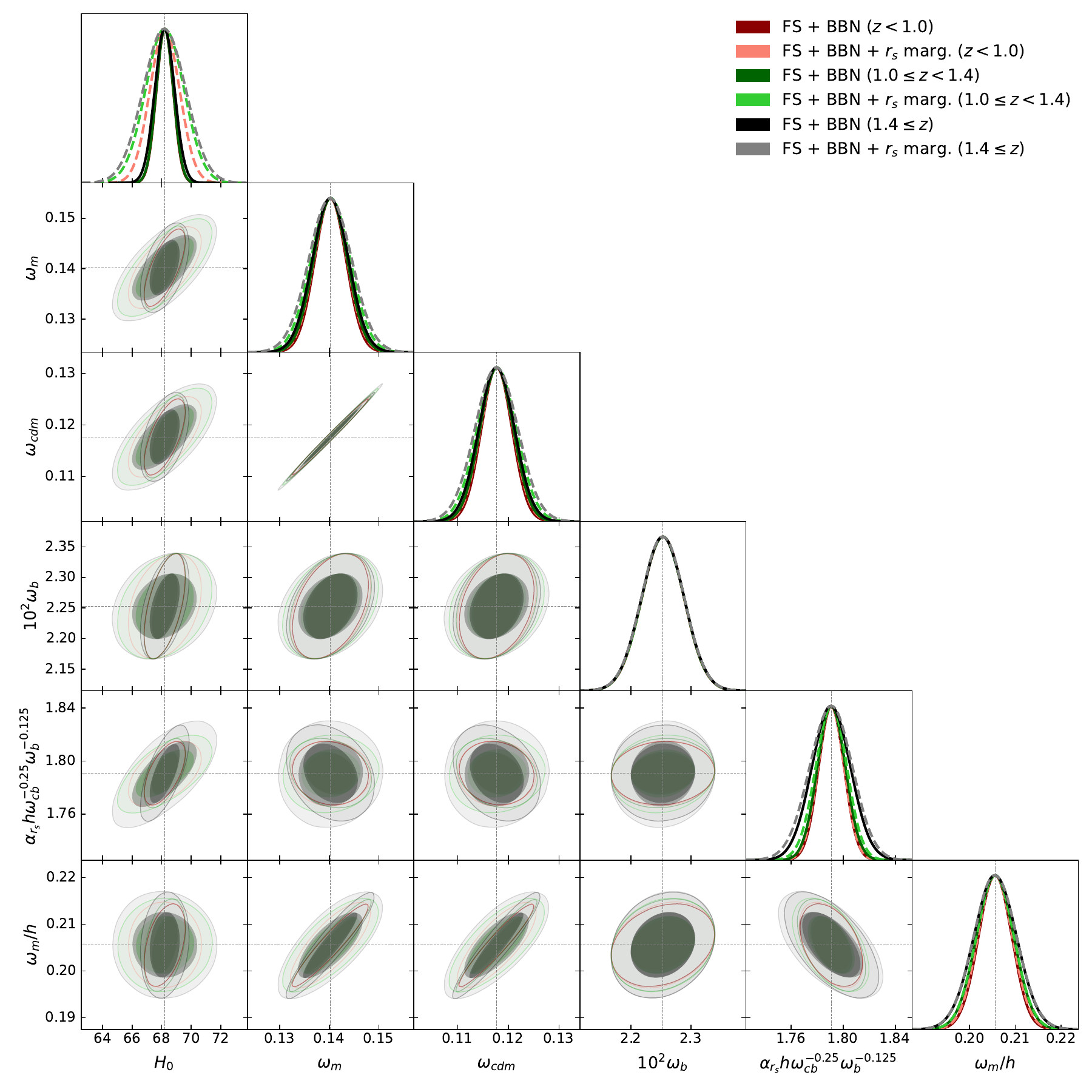}
    \caption{\resub{Fisher forecast of $\Lambda$CDM parameter constraints for a a Euclid-like survey, subdividing the data into low-, medium- and high-redshift subsets as described in Appendix \ref{app:kmax_z_vary}. The marginalised constraints on $H_0$ are given in Table~\ref{tab:kmax_z_forecast}. We observe that $r_s$-marginalisation more significantly degrades constraints from the medium- and high-redshift datasets increasing the relative weight of the low-redshift data in our analysis compared to a non-$r_s$-marginalised analysis.}}
    \label{fig:z_forecast_plot}
\end{figure}

\resub{Secondly, we divide our data into low-, medium- and high-redshift bins. As discussed above, the fiducial analysis uses eight redshift bins evenly spaced between $z=0.5$ and $z=2.1$. Our low-redshift dataset contains the lowest two redshift bins ($\bar{z}=0.6$ and $0.8$), the medium-redshift sample consists of the redshift bins with mean redshifts $\bar{z}=1.0$ and $1.2$, while the remaining four redshift bins ($\bar{z}=1.4$, $1.6$, $1.8$ and $2.0$) make up the high-redshift subset. One can see in Fig.\,\ref{fig:z_forecast_plot} that without $r_s$-marginalisation, we expect similar $H_0$ constraints from each of the three subsets (see Table~\ref{tab:kmax_z_forecast} for forecast uncertainties). When including our sound horizon marginalisation procedure, the high- and medium-redshift constraints are more significantly degraded. Our analysis thus increases the relative weight of the low-redshift data.}

\begin{table}[]
    \caption{\resub{$H_0$ constraints from $\Lambda$CDM fisher forecast for a Euclid-like survey, using various choices of $k_{\rm max}$ (left) and redshift binning (right), as described in Appendix \ref{app:kmax_z_vary}.}\label{tab:kmax_z_forecast}}
    \centering
    \begin{tabular}{d|d|d}
    \hline\hline
     &\multicolumn{2}{c}{$\sigma_{H_0}$ $[\hun]$}\\
     \multicolumn{1}{c|}{$k_{\rm{max}}$ $[\hMpc]$} & \multicolumn{1}{c|}{\quad FS + BBN \quad} & \multicolumn{1}{c}{\quad FS + BBN + $r_s$ marg.} \\\hline
     0.1 & 1.3 & 1.9\\
     0.5 & 0.44 & 0.86\\
     1.0 & 0.43 & 0.72\\
     \hline\hline
    \end{tabular}\quad
    \begin{tabular}{c|d|d}
    \hline\hline
     &\multicolumn{2}{c}{$\sigma_{H_0}$ $[\hun]$}\\
     \multicolumn{1}{c|}{$z$ selection} & \multicolumn{1}{c|}{\quad FS + BBN \quad} & \multicolumn{1}{c}{\quad FS + BBN + $r_s$ marg.} \\\hline
     $\bar{z}<1.0$ & 0.57 & 1.0\\
     $1.0\leq \bar{z} < 1.4$ & 0.58 & 1.3\\
     $1.4 \leq \bar{z}$ & 0.67 & 1.4\\
     \hline\hline
    \end{tabular}
\end{table}

\bibliography{bibliography1_Philcox++,bibliography2_Philcox++,bibliography_H0_wo_rd_II}

\end{document}